\documentclass[fleqn,usenatbib]{mnras}

\usepackage{newtxtext,newtxmath}

\usepackage[T1]{fontenc}
\usepackage{ae,aecompl}

\usepackage{graphicx}	

\usepackage{amsmath}	
\usepackage{amssymb}	
\usepackage{comment}
\usepackage{siunitx}
\usepackage{booktabs}
\usepackage{gensymb}
\usepackage{threeparttable} 

\usepackage{xcolor}

\let\oldnabla\nabla
\renewcommand{\nabla}{\vec{\oldnabla}}

\def\be{\begin{equation}}
\def\ee{\end{equation}}
\newcommand\code[1]{\textsc{\MakeLowercase{#1}}}

\def\gsim{\lower.5ex\hbox{\gtsima}} 
\def\lsim{\lower.5ex\hbox{\ltsima}} 
\def\gtsima{$\; \buildrel > \over \sim \;$} 
\def\ltsima{$\; \buildrel < \over \sim \;$} \def\gsim{\lower.5ex\hbox{\gtsima}} 
\def\lsim{\lower.5ex\hbox{\ltsima}} 
\def\simgt{\lower.5ex\hbox{\gtsima}} 
\def\simlt{\lower.5ex\hbox{\ltsima}}

\def\msun{{\rm M}_{\odot}}
\def\lsun{{\rm L}_{\odot}}

\def\S*{$\Sigma_{\rm SFR}$}

\def\AU{\mathrm{AU}}

\graphicspath{{plots/}} 

\usepackage{newtxtext,newtxmath}

\usepackage[T1]{fontenc}

\DeclareRobustCommand{\VAN}[3]{#2}
\let\VANthebibliography\thebibliography
\def\thebibliography{\DeclareRobustCommand{\VAN}[3]{##3}\VANthebibliography}

\usepackage{graphicx}	
\usepackage{amsmath}	

\newcommand{\alphaSS}{\alpha_{\rm{SS}}}

\title[Turbulence of DSHARP discs]{Constraining turbulence in protoplanetary discs using the gap contrast: an application to the DSHARP sample}

\author[Pizzati et al.]{Elia Pizzati$^{1}$\thanks{\href{mailto:pizzati@strw.leidenuniv.nl}{pizzati@strw.leidenuniv.nl}},
Giovanni P. Rosotti$^{1,2,3}$,
Benoît Tabone$^{4}$
\\
$^{1}$ Leiden Observatory, Leiden University, P.O. Box 9513, 2300 RA Leiden,
The Netherlands\\
$^{2}$ School of Physics and Astronomy, University of Leicester, Leicester LE1 7RH, UK\\
$^{3}$ Dipartimento di Fisica 'Aldo Pontremoli', Università degli Studi di Milano, via G. Celoria 16, I-20133 Milano, Italy\\
$^{4}$ Université Paris-Saclay, CNRS, Institut d’Astrophysique Spatiale, 91405 Orsay, France\\
}

\date{Accepted XXX. Received YYY; in original form ZZZ}

\pubyear{2023}

\begin{document}
\label{firstpage}
\pagerange{\pageref{firstpage}--\pageref{lastpage}}
\maketitle

\begin{abstract}
Constraining the strength of gas turbulence in protoplanetary discs is an open problem that has relevant implications for the physics of gas accretion and planet formation. In this work, we gauge the amount of turbulence in 6 of the discs observed in the DSHARP programme by indirectly measuring the vertical distribution of their dust component. We employ the differences in the gap contrasts observed along the major and the minor axes due to projection effects, and build a radiative transfer model to reproduce these features for different values of the dust scale heights. We find that (a) the scale heights that yield a better agreement with data are generally low ($\lesssim4\,\AU$ at a radial distance of $100\,\AU$), and in almost all cases we are only able to place upper limits on their exact values; these conclusions imply (assuming an average Stokes number of $\approx10^{-2}$) low turbulence levels of $\alphaSS\lesssim10^{-3}-10^{-4}$; (b) for the 9 other systems we considered out of the DSHARP sample, our method yields no significant constraints on the disc vertical structure; we conclude that this is because these discs have either a low inclination or gaps that are not deep enough. Based on our analysis we provide an empirical criterion to assess whether a given disc is suitable to measure the vertical scale height.
\end{abstract}




\section{Introduction} \label{sec:intro}

Characterising the magnitude of turbulence in accretion discs is a classical problem in astrophysics. This is because turbulence is often invoked (see historical discussion in \citealt{Pringle1981}) as the mechanism responsible for powering accretion. On the one hand, therefore, the first scientific question that any such study seeks to address is whether the level of turbulence, commonly quantified through the $\alphaSS$ parameter \citep{ShakuraSunyaev1973}, is high enough to explain the observed accretion rates. For the specific case of proto-planetary discs we study in this paper, this is a particularly important question: the cold conditions of these discs, which are clearly in the non-ideal magnetohydrodynamics regime, make it far from obvious to understand whether the magneto-rotational instability \citep{BalbusHawley} can be a mechanism responsible for generating the required level of turbulence. Addressing this question, and studying in parallel other processes that could generate turbulence in proto-planetary discs, is a subject of many studies (see \citealt{LesurPPVII} for a recent review).

For proto-planetary discs, the issue runs even deeper than the question about accretion; even if turbulence was ultimately found not to be responsible for accretion, it would still affect a wealth of processes and therefore have a strong impact on planet formation. A non-exhaustive list of processes affected by turbulence includes the heating and cooling balance in the terrestrial planet-forming region due to the importance of viscous heating \citep{MinDullemond2011}, the diffusion of molecular species radially \citep{Owen2014Diffusion} and vertically \citep{SemenovWiebe2011,Krijt2020}, the diffusion of dust, setting both the dust disc vertical extent \citep{Dubrulle1995} and the leakiness of dust traps (e.g., \citealt{Zormpas2022}). For what concerns planets in particular, turbulence has a profound impact on disc-planet interaction; its magnitude affects \citep{PaardekooperPPVII} the ability of planets to open gaps in the disc and how fast they migrate by exchanging angular momentum with the disc. Turbulence is also a crucial parameter setting how quickly planets accrete gas \citep{Bodenheimer2013} and dust \citep{JohansenLambrechtsReview} from the disc, determining the final masses of planetary systems. Last but not least, turbulence controls the onset of the streaming instability \citep{DrazkowskaPPVII}, one of the best known mechanisms for creating planetesimals and kick-starting the planet formation process.

It would thus be beneficial to have a method to constrain turbulence observationally. Thankfully, in the last few years, the field has been completely transformed by the Atacama Large Millimeter Array (ALMA), which provided order-of-magnitude improvements in sensitivity and angular resolution. First of all, by studying line broadening of emission lines, ALMA has allowed to directly detect turbulence in two discs \citep{Flaherty2020,FlahertyPrep}, and only yielding upper limits in a limited number of other cases \citep{PintePPVII}. In addition, ALMA has opened up many other observational routes (recently reviewed in \citealt{RosottiTurbulenceReview}) for indirectly constraining turbulence. These routes include the study of the disc vertical thickness, the radial extent of dust and gas rings, and population studies that use disc demographics studies (see \citealt{ManaraPPVII} for a review), that is, catalogues of the fundamental disc properties (such as mass, radius and mass accretion rate) for large disc samples. In this way, in the last few years the study of disc turbulence has moved from an almost theoretical subject to an observational one.

In this paper, we constrain turbulence by measuring the disc vertical thickness. The vertical equilibrium of dust grains is a competition between settling, which is determined by the joint action of gas drag and gravity, and turbulence, which stirs up the grains in the vertical direction. In simple terms, then, the more turbulent the disc is, the thicker it is, but it should be highlighted that the presence of drag implies that the aerodynamic coupling between gas and dust (normally parametrized by the Stokes number $\mathrm{St}$) also influences the thickness. Indeed, as we will recap in Sec. \ref{sec:discussion}, the method is only sensitive to $\alphaSS/\mathrm{St}$.

\begin{figure}
	\centering
	\includegraphics[width=0.45\textwidth]{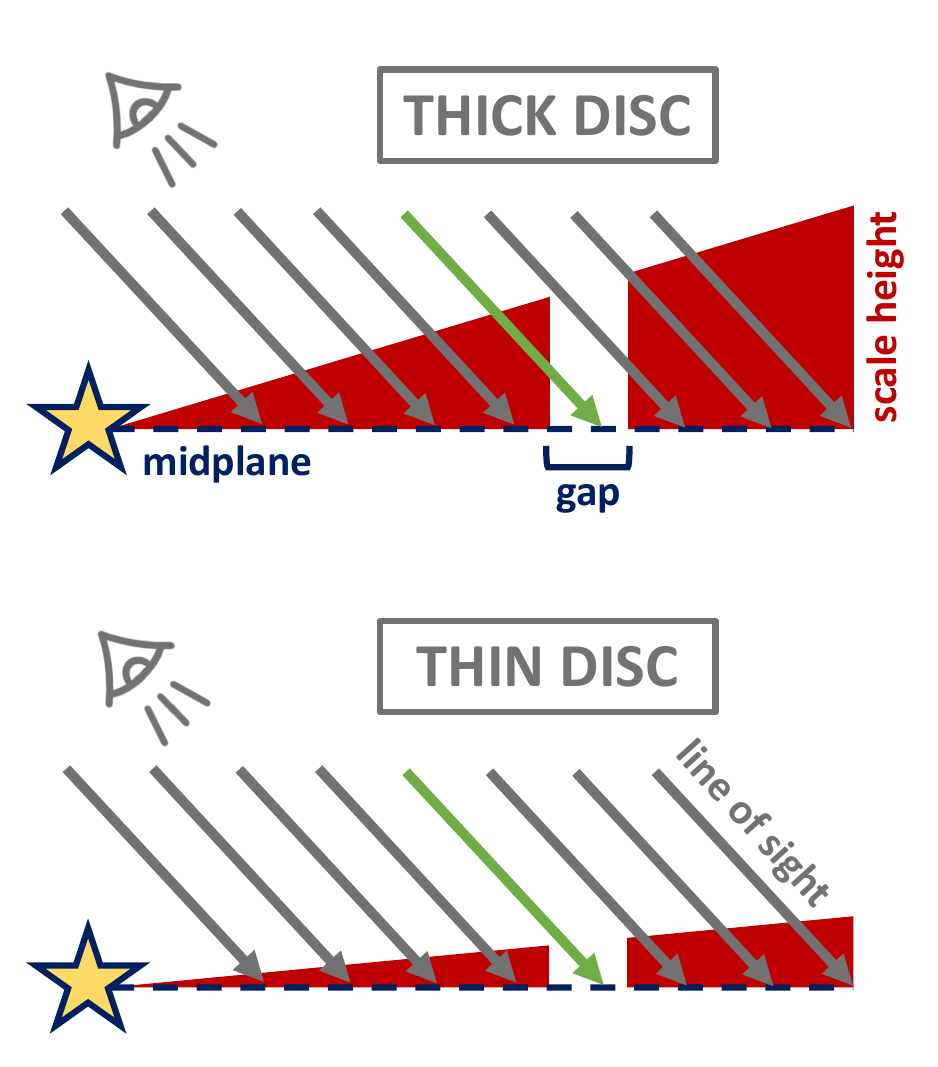}
    \caption{Sketch of the \textit{gap-filling effect}. The lines of sight intercept the disc's plane with an angle that depends on the disc inclination. In the presence of a gap, lines of sights piercing through the gap (e.g., the one highlighted in green in the sketch) may still intercept sections of the disc that are far from the mid-plane due to simple geometrical effects. Therefore, the gap will be seen as partially filled by the observer. If the disc is thicker (thinner), this filling effect is stronger (weaker); this implies that the \textit{gap-filling effect} can be used to indirectly gauge the vertical extension of the disc. \label{fig:intro_proj_effect}
 	}
\end{figure}

We apply the technique developed by \citet{Pinte2016} in their study of HL Tau. The technique relies on the fact that many observed discs \citep{BaePPVII} present an emission pattern characterised by bright rings and dark gaps. \citet{Pinte2016} realised that due to projection effects in a disc with finite thickness the line of sight will intercept sections of the disc that are out of the midplane. In a gap, this has the effect that the adjacent bright regions partially contaminate the dark gap, lowering the gap depth. We will refer to this in the rest of the paper as the \textit{gap-filling effect}. It is easy to realise that the geometry of projection is such that this filling effect is much larger along the minor axis of the disc than along the major axis. Once the image is deprojected in polar coordinates, as commonly done in the field, the resulting effect is that the gaps are more ``filled'' (i.e., shallower) along the disc minor axis and more ``empty'' (i.e., deeper) along the disc major axis. The difference between minor and major axes increases with the disc thickness and therefore it is a way to probe the vertical structure of the disc. A simple sketch in Figure \ref{fig:intro_proj_effect} shows the simple geometrical argument behind the \textit{gap-filling effect}. Extracting quantitative measurements from this effect requires building radiative transfer models of the emission. 

So far, in addition to HL Tau, the method has been applied to HD163296 \citep{Liu2022HD163296} and to Oph163131 \citep{Villenave2022}. The goal of this paper is to significantly expand this observational sample. For this purpose we selected the sample of the Large Programme DSHARP \citep{Andrews2018}, which consists of twenty discs imaged at 0.05" resolution, since it constitutes the largest homogeneous high-resolution survey of proto-planetary discs. We aim to determine in which cases this technique is successful in gauging the
disc thickness, and, whenever possible, to place meaningful constraints on the disc scale heights.

This paper is structured as follows: in Sec. \ref{sec:methods}, we discuss the basic assumptions of our model and describe the steps of the analysis we perform to match DSHARP data. A description of the data sample is presented in Sec. \ref{sec:data}. Sec. \ref{sec:results} presents the main results of the analysis, while Sec. \ref{sec:discussion} contains a discussion on the implications of our findings. Conclusions are given in Sec. \ref{sec:conclusions}.

\section{Methods} \label{sec:methods}

In this section, we describe the basic assumptions of the model and provide details on the methodology we employ to compare our synthetic images with DSHARP observations. 

\subsection{Disc structure} \label{sec:disc_structure}

In the following, we are only interested in modelling the dust component of the protoplanetary discs, as the ALMA observations we use are only focusing on the dust continuum emission. Therefore, in this section -- whenever not stated otherwise --  we refer with the term ``disc'' to the dust component only. 
In Sec. \ref{sec:discussion}, we will discuss further how our results can be employed to study the relationship between dust and gas in the disc, and ultimately to constrain the amount of gas turbulence. 

We model the disc as a cylindrically-symmetric system with a minimum radius $r_{\rm in }$ and a maximum radius $r_{\rm out }$. We assume for the vertical density distribution a Gaussian profile of the form:
\begin{equation}
    \rho_d(r,z) = \frac{\Sigma(r)}{\sqrt{2\pi}h_d(r)} \,\exp\left(-\frac{z^2}{2h_d(r)^2}\right),
    \label{eq:dust_profile}
\end{equation}
where $\Sigma(r)$ is the dust surface density and $h_d(r)$ is the dust scale height. 

This profile originates from an analogy to the gas component, which can be assumed to be in hydrostatic equilibrium along the vertical direction and thus follows a Gaussian profile identical to eq. \ref{eq:dust_profile}. Formally, the dust has a different equilibrium solution, but eq. \ref{eq:dust_profile} is a close approximation \citep{Dubrulle1995}. Furthermore, defining the dust density in this way is convenient as we can consider the ratio between gas and dust scale-heights, which will be important to estimate turbulence.

In Sec. \ref{sec:iteration}, we discuss our procedure to determine the disc surface density $\Sigma(r)$. As for the dust scale height profile, we assume a simple flaring model:
\begin{equation}
    h_d(r) = H_d\left(\frac{r}{r_0}\right)^{1.25},
    \label{eq:dust_scaleheight_radial}
\end{equation}
In what follows, we set the reference radius $r_0$ to $100\,\AU$ and take the value of the dust scale height at this radius, $H_d=h_d(r=100\,\AU)$, as the only free parameter of our model. The goal of our work is to gauge the value of $H_d$ using the \textit{\textit{gap-filling effect}} on the disc minor axis. 

In order to predict the observed surface brightness of the disc, we need to determine the dust temperature profile. For this, we assume that the dust is passively heated by the central star, and that a fraction $\phi_{\rm flux}$ of the total flux emitted by the star is intercepted by the disc. Following radiative-transfer models \citep[e.g.,][]{ChiangGoldreich1997,D'Alessio1998,Dullemond2001,Dullemond2018}, we set $\phi_{\rm flux}=0.02$ and write ($\sigma$ is the Stefan-Boltzmann constant):
\begin{equation}
T(r) = \left(\frac{0.5\phi_{\rm flux}L_\star}{4\pi r^2\sigma}\right)^{1/4} = T_{\rm in}\,\left(\frac{r_{\rm in}}{r}\right)^{0.5},
\label{eq:temperature}
\end{equation}
where we have expressed everything in terms of the temperature at the inner radius, $T_{\rm in }$. For the sake of simplicity, we use this analytical description of the disc temperature for our model -- instead of self-consistently computing the temperature using a Montecarlo approach \citep[see e.g.,][]{Liu2022HD163296}. In Sec. \ref{sec:caveat}, we discuss the reasons for such choice and the caveats that come with it.

The expected surface brightness of the disc can then be determined assuming dust thermal emission. In order to create mock images of our disc models, we use the code \code{radmc-3d}\footnote{\url{http://www.ita.uni-heidelberg.de/~dullemond/software/radmc-3d/}}. We set the extrinsic parameters (such as the distance, sky coordinates, inclination, position angle) in accordance with observations (see Sec. \ref{sec:data}), and we produce synthetic images of the discs according to the radiative transfer computation. Then, we use the \code{CASA} package \citep{CASAPaper} to produce mock observations with the same beam and antennae configuration of the original ALMA data. In order to do this we have retrieved from the DSHARP Data Release webpage\footnote{\url{https://almascience.eso.org/almadata/lp/DSHARP/}} the visibility files of the DSHARP observations. In our analysis we use the same version of \code{CASA} used by the DSHARP team (v 5.1.1-5) to ensure that the data and the models have been processed in the same way\footnote{That being said, for safety we have recomputed the CLEAN images for the data starting from the visibility files, in order to be sure that we use the same CLEAN parameters in the data as in the models.}. We created synthetic visibilities from the radiative transfer image at the uv coordinates of the observations using the \code{CASA} task ft. We then apply the \code{CLEAN} algorithm to generate a synthetic ALMA image to compare with the observed image. We use the scripts provided by the DSHARP team in order to make sure that we use the same \code{CLEAN} parameters as the observations. To reduce the computational time, it is common in the field to employ the simpler approach of a convolution with a Gaussian beam. While this is often satisfactory, we noticed in early tests that the detailed shape of the emission profile in the gap is different from images produced by the \code{CLEAN} algorithm. In addition, some of the DSHARP sources have clear \code{CLEAN} artefacts such as negative emission that cannot be reproduced with a simple Gaussian convolution. Therefore, we adopt here a consistent approach to include the contribution of these cleaning artifacts.

\subsection{Inferring the disc surface density} \label{sec:iteration}

\begin{figure}
	\centering
	\includegraphics[width=0.52\textwidth]{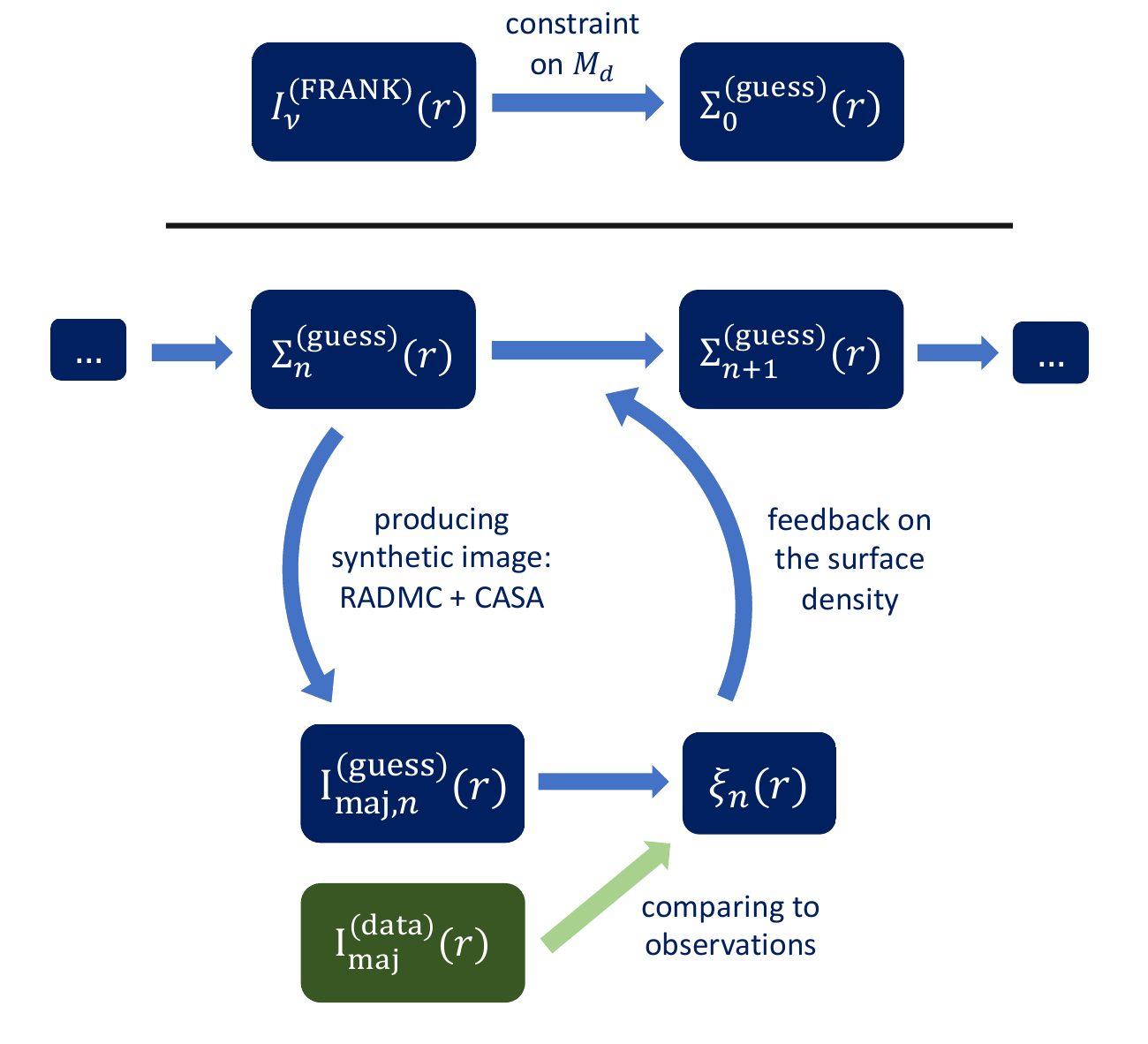}
    \caption{Overview of the iteration procedure we employ to extract the disc surface density, $\Sigma(r)$. As described in Sec. \ref{sec:iteration}, we choose the profile for the surface density that matches the observed intenstity profile along the major axis. The various quantities appearing in the sketch are defined in the main text. The top part of the sketch refers to the kickstarting of the process (where we find a first guess for the surface density profile), while the bottom part shows an instance of a single iteration.  
    \label{fig:iteration}
 	}
\end{figure}

In order to proceed further with our analysis, we need to infer the surface density $\Sigma(r)$ of the observed disc. This is not straightforward, as simple power-law models are not capable to reproduce the wealth of substructures (gaps and rings) that are observed in the DSHARP images. Given that our goal is to use the \textit{gap-filling effect} as a probe of the disc vertical size, modeling these substructures within a reliable framework is of paramount importance. Therefore, similarly to what was done in \citet{Pinte2016}, we employ here an iterative procedure to find the correct surface density of our discs. We outline the procedure in the following paragraphs, and provide an overview of the different steps involved in the iteration cycle in Figure \ref{fig:iteration}.

The fundamental idea we adopt in this procedure is that the intensity observed along the major axis is a good proxy for the real surface density of the disc. This is because, as already discussed in Sec. \ref{sec:intro}, the \textit{gap-filling effect} affects only marginally the major axis, whereas it has the strongest effect on the minor axis. Therefore, our goal is to find via multiple iterations a surface density profile that is able to match the intensity observed along the major axis. 

The procedure can be summarized as follows. First of all, we need to extract the intensity along the major axis, $I_{\rm maj}^{(\rm data)}(r)$, from the 2D images. In order to do that, for every disc we analyze, deproject the image and average two opposite slices of $1/8$ (i.e., with a width of $\pi/4$) of the disc centered on the major axis. When deprojecting the disc emission maps, we make sure that the images are aligned with the discs' centres by using the offsets in the $x$ and $y$ coordinates reported by \citet{DSHARPII} (see their Tab. 2). The resulting $I_{\rm maj}^{(\rm data)}(r)$ represents our benchmark profile that we aim to reproduce with a suitable choice of the disc surface density.

Then, we use as a first guess for the surface density profile the output of the \code{Frankenstein} \citep{FrankesteinPaper} fit of the DSHARP sources presented in \citet{JenningsDSHARP}. \code{Frankenstein} is a code that uses Gaussian processing to fit disc emission profiles in visibility space, using the assumption that the emission is azimuthally symmetric. This gives a good starting point for the initial surface density since \code{Frankenstein} can achieve a spatial resolution higher than the \code{CLEAN}ed images we analyse in this paper. While this gives us the shape of the surface density, note that we also need a normalisation constant: Frankestein fits for the emission profile (giving a profile $I_\nu^\mathrm{(FRANK)}(r)$ as an output), while we need a surface density to give as input to \code{radmc-3d}. In order to convert the intensity profile into a surface density, we use as a constraint the formula often employed \citep{BeckwithSargent1990} to estimate the disc mass $M_\mathrm{dust}$ from sub-mm observations:
\begin{equation}
M_\mathrm{dust} =  \frac{F_\nu d^2}{\kappa_\nu B_\nu (T_\mathrm{dust})}
\end{equation}
where $F_\nu$ is the flux in the image, $d$ the distance to the source; for $T_\mathrm{dust}$ we take a temperature of 20 K and $\kappa_\nu$ is the opacity of the dust we employ. Since we consider a single grain population, the physical quantity we are constraining is the dust optical depth (given the prescribed temperature profile), and not the dust surface density. This implies that the value of the opacity only acts as a normalisation for the dust surface density and does not have any influence on our conclusions - with a different dust opacity we would simply need to change the dust surface density accordingly in order to have the same optical depth. Notice also that the formula is only an approximation (the emission is not guaranteed to be optically thin and 20 K may not be the correct value); however the value reported above is only needed to kickstart the iteration and the iterative procedure will take care of reaching the correct values, both for the normalisation and for the shape of the surface density. Finally, because we fitted for the emission profile but need the surface density, we multiply the resulting profile by $r^{1/2}$ to take into account the variation in disc temperature with radius\footnote{In the same way as for the normalisation, this is only a first order correction; the iterative procedure will better refine this radial scaling.}. Let us call $\Sigma_0^{(\rm guess)}(r)$ the surface density we have obtained in this way. We use this guess to define our disc structure (setting also the dust scale height parameter $H_d$ to a fixed value) and produce synthetic observations using \code{radmc3d} + \code{CASA} (see Sec. \ref{sec:disc_structure}). 

Subsequently, we apply the same procedure as described above to deproject these mock observations and to extract a mock intensity profile along the major axis\footnote{With the exception of AS 209, where the procedure described here uses the azimuthally averaged intensity profile rather than the major axis (see also \autoref{tab:sample}).}, $I_{\rm maj,0}^{(\rm guess)}(r)$. This profile can be directly compared to the observational one, $I_{\rm maj}^{(\rm data)}(r)$. This comparison outputs a ratio, $\xi_0(r)=I_{\rm maj}^{(\rm data)}/I_{\rm maj,0}^{(\rm guess)}$ that parametrizes how well the initial guess for the surface density is able to reproduce observations. We can improve this match simply by multiplying the initial guess for the surface density profile, $\Sigma_0^{(\rm guess)}(r)$, for the ratio $\xi_0(r)$, finding a new guess for the disc surface density $\Sigma_1^{(\rm guess)}(r)$. To prevent large variations of the surface density from one iteration to the next, we do not allow variations larger than a factor 4 in a single iteration. We then iterate this procedure by using this new surface density profile to produce mock observations $I_{\rm maj,1}^{(\rm guess)}$ and update the surface density using the sequence:
\begin{equation}
    \Sigma_{n+1}^{(\rm guess)}(r) = \xi_n(r) \Sigma_{n}^{(\rm guess)}(r) = \frac{I_{\rm maj}^{(\rm data)}}{I_{\rm maj,n}^{(\rm guess)}}\Sigma_{n}^{(\rm guess)}(r)
\end{equation}

We stop the iteration when a value of $|\xi(r) -1|<0.05$ is reached for every radius $r$. On average, this takes around $10-15$ cycles. As expected, the convergence is very easily achieved where the intensity profile is smooth, whereas it takes more iterations in the regions where gaps and rings are present, especially when they are narrow and deep. 
For a few systems, this implies that convergence is not reached at the bottom of the deepest gaps even after $15$ iterations, with the difference between the model and data being in the range $5-10\%$. We empirically find that increasing the number of iterations does not give any significant advantage for these peculiar systems, with only minimal gains in terms of model-data accordance despite the large number of iterations employed. Therefore, we decide to set a maximum iteration number $n=15$ and insert a caveat for the systems that are not converged inside the gaps according to our criteria (Sec. \ref{sec:data}).

\subsection{Analysis of the gaps filling effect on the minor axis}

At the end of the iteration cycle described in Sec. \ref{sec:iteration}, we obtain, for a fixed value of dust scale height $H_d$, a fiducial profile for the disc surface density, $\Sigma^{(\rm model)}(r; H_d)$. Note that the dependence of $\Sigma^{(\rm model)}(r; H_d)$ on $H_d$ is very mild, as the surface density is obtained by comparing the model with the data along the major axis, where the vertical structure of the disc has only a small effect on the final intensity. This $\Sigma^{(\rm model)}(r; H_d)$ corresponds to an intensity profile along the major axis -- $I_{\rm maj}^{(\rm model)}(r;H_d)$ -- that matches the observed one -- $I_{\rm maj}^{(\rm data)}(r)$ (see also the sketch in Fig. \ref{fig:iteration}).

Our goal is then to extract, both from the observational data and from our fiducial model, the intensity profiles along the minor axis -- $I_{\rm min}^{(\rm data)}(r)$ and $I_{\rm min}^{(\rm model)}(r;H_d)$, respectively. Similarly to what described in Sec. \ref{sec:iteration}, in order to do this, we take the deprojected 2D images and average two opposite slices of 1/8 of the disc centered on the minor axis.

Along the minor axis, the predicted intensity can depend quite strongly on the value of the dust scale height $H_d$, as the vertical thickness of the disc directly influences the gap filling along the minor axis. Therefore, a simple comparison between $I_{\rm min}^{(\rm data)}(r)$ and $I_{\rm min}^{(\rm model)}(r;H_d)$ offers a way to place constraints on the vertical structure of the disc. A quantitative analysis of this comparison and the implications of the results we find are presented in Sec. \ref{sec:results} and \ref{sec:discussion}.

\section{DSHARP data sample} \label{sec:data}

In this section, we describe the systems we use for our analysis. DSHARP is a very high resolution ($\sim$0.035", or 5 AU) observational campaign that targeted 20 proto-planetary discs with the goal of finding and characterising substructures in the dust continuum emission at 240 GHz. We examined the entire DSHARP catalogue and excluded the systems that are not suitable for our study of the \textit{gap-filling effect}. These include 3 single systems that show signs of spirals (i.e, IM Lup, Elias 27, and WaOph 6) and two binary systems (HT Lup and AS 205), where the individual discs either show signs of spirals or lack clear substructure. Spirals are not compatible with the assumption of perfect azimuthal symmetry in our disc model and we therefore discard the discs showing this signature. 

For the remaining 15 systems, we run our model to find the best matching value(s) of the dust scale height $H_d$. We describe the results of this analysis in the following section. Here, we provide more details on the properties of these systems. In Table \ref{tab:sample}, we report the parameters of the systems as listed by \citet{Andrews2018}: these include the mass and luminosity of the central star, the distance of the system, the inclination angle, the position angle (PA), the outer radius of the disc, and the beam size. The inner radius cannot be determined easily from observations, and thus we always set it to $r_{\rm in} = 2\, \AU$. This choice has no relevant impact on the final results since we are only interested here in radial locations with gaps.

\begin{table*}
\centering
\setlength{\extrarowheight}{3pt}
\begin{threeparttable}

\begin{tabular}{ c | c c c c c c c  | c c c }
\toprule
\multicolumn{1}{c}{}& \multicolumn{7}{c}{{Observational data}}
& \multicolumn{3}{c}{{Model Predictions}}\\
 ID  & $\log M_\star$    & $\log L_\star$   & $D$   & $i$     & PA & $r_{\rm out}$  & beam & $H_d$ & 
 $\chi^2_{\rm norm}$ & 
 Constraint\\ 

   & [$\msun$] &  [$\lsun$]   & [pc]    &  [$\degree$]     &  [$\degree$] & [$\AU$] & [ $"\times"$ ] & [$\AU$] &  &on $H_d$\\ 
 \midrule
\midrule
GW Lup    &    
$-0.34^{+0.10}_{-0.17}$    &   
$-0.48\pm0.20$    &   
$155\pm 3$   &   
$38.7\pm0.3$    & 
$37.6\pm0.5$  &
$105$   &   
$0.045\times0.043$   & 
$\lesssim 4$ &
$0.27$&
Upper limit\\
DoAr 25    &    
$-0.02^{+0.04}_{-0.19}$    &   
$-0.02\pm0.20$    &   
$138\pm 3$   &   
$67.4\pm0.2$    & 
$110.6\pm0.2$  &
$165$   &   
$0.041\times0.022$   & 
$\lesssim 2$ &
$4.5$&
Upper limit\\
Elias 24 \tnote{a}   &    
$-0.11^{+0.16}_{-0.08}$    &   
$0.78\pm0.20$    &   
$136\pm 3$   &   
$29.0\pm0.3$    & 
$45.7\pm0.7$  &
$160$   &   
$0.037\times0.034$   & 
$\lesssim 2$ &
$4.9$&
Upper limit\\
HD 142666    &    
$0.20^{+0.04}_{-0.01}$    &   
$0.96\pm0.21$    &   
$148\pm2 $   &   
$62.22\pm0.14$    & 
$162.11\pm0.15$  &
$80$   &   
$0.032\times0.022$   & 
$-$ &
$-$&
No constraint\\
AS 209  \tnote{c}  &    
$-0.08^{+0.11}_{-0.14}$    &   
$0.15\pm0.20$    &   
$121\pm 2$   &   
$34.97\pm0.13$    & 
$85.76\pm0.16$  &
$160$   &   
$0.038\times0.036$   & 
$\lesssim 2$ &
$0.79$&
Upper limit\\
Elias 20    &    
$-0.32^{+0.12}_{-0.07}$    &   
$0.35\pm0.20$    &   
$138\pm 5$   &   
$49\pm1$    & 
$153.2\pm1.3$  &
$85$   &   
$0.032\times0.023$   &  
$-$ &
$-$&
No constraint\\
Sz 129    &    
$-0.08^{+0.03}_{-0.15}$    &   
$-0.36\pm0.20$    &   
$161\pm 3$   &   
$34.1\pm1.3$    & 
$151\pm2$  &
$95$   &   
$0.044\times0.031$   & 
$-$ &
$-$&
No constraint\\
HD 163296 \tnote{b}   &    
$0.31^{+0.05}_{-0.03}$    &   
$1.23\pm0.30$    &   
$101\pm 2$   &   
$46.7\pm0.1$    & 
$133.33\pm0.15$  &
$170$   &   
$0.048\times0.038$   & 
$\approx 4$ &
$25$&
Upper/lower lim.\\
HD 143006    &    
$0.25^{+0.05}_{-0.08}$    &   
$0.58\pm0.15$    &   
$165\pm 5$   &   
$18.6\pm0.8$    & 
$169\pm2$  &
$105$   &   
$0.046\times0.045$   & 
$-$ &
$-$&
No constraint\\
SR 4 \tnote{a}   &    
$-0.17^{+0.11}_{-0.14}$    &   
$0.07\pm0.20$    &   
$134\pm 2$   &   
$22\pm2$    & 
$18\pm5$  &
$85$   &   
$0.034\times0.034$   & 
$-$ &
$-$&
No constraint\\
RU Lup    &    
$-0.20^{+0.12}_{-0.11}$    &   
$0.16\pm0.20$    &   
$159\pm 3$   &   
$18.8\pm1.6$    & 
$121\pm5$  &
$80$   &   
$0.025\times0.024$   & 
$-$ &
$-$&
No constraint\\
MY Lup    &    
$0.09^{+0.03}_{-0.13}$    &   
$-0.06\pm0.20$    &   
$156\pm 3$   &   
$73.2\pm0.1$    & 
$58.8\pm0.1$  &
$115$   &   
$0.044\times0.043$   & 
$\lesssim 4$ &
$0.58$&
Upper limit\\
Sz 114    &    
$-0.76^{+0.08}_{-0.07}$    &   
$-0.69\pm0.20$    &   
$162\pm 3$   &   
$21.3\pm1.3$    & 
$165\pm4$  &
$65$   &   
$0.067\times0.028$   & 
$-$ &
$-$&
No constraint\\
WSB 52    &    
$-0.32^{+0.13}_{-0.17}$    &   
$-0.15\pm0.20$    &   
$136\pm 3$   &   
$54.4\pm0.3$    & 
$138.4\pm0.3$  &
$39$   &   
$0.033\times0.027$   & 
$-$ &
$-$&
No constraint\\
DoAr 33    &    
$0.04^{+0.05}_{-0.17}$    &   
$-0.18\pm0.20$    &   
$139\pm 2$   &   
$41.8\pm0.8$    & 
$81.1\pm1.2$  &
$27$   &   
$0.037\times0.024$   & 
$-$ &
$-$&
No constraint\\
 \bottomrule
\end{tabular}
 \begin{tablenotes}
       \item [a] Convergence (described in Sec. \ref{sec:iteration}) cannot be reached. The (maximum) relative difference between the model and the data does not go below the threshold value of $5\%$; instead, it sits between $5\%$ and $10\%$.
       \item [b] When determining the intensity along the major axis, we use only one side of the major axis since the other one presents a feature that disrupts azimuthal symmetry.
       \item[c] For this system, in order to determine the surface density via our convergence procedure, we do not consider the intensity along the major axis but the average. This is because the former presents some negative values in the gaps and this creates an issue with the convergence procedure described in Sec. \ref{sec:iteration}.
     \end{tablenotes}
     \caption{Properties of the DSHARP systems considered here for the analysis. Observational parameters are taken from \citet{Andrews2018} and \citet{DSHARPII}. From left to right: name of the source; mass and luminosity of the central star ($M_\star$, $L_\star$), distance ($D$), inclination angle ($i$), position angle (PA), maximum radius ($r_{\rm out}$), beam size (beam); results from our analysis, i.e., the best-fitting value of $H_d$ and the associated reduced chi-squared, $\chi^2_{\rm norm}$. The final column describes whether it was possible to use our analysis to constrain the scale height in a meaningful range. The units of measurements for every parameters are shown on the second row; uncertainties for the disc extensions and the beam size are irrelevant to our discussion and not shown here. 
}
     \label{tab:sample}
  \end{threeparttable}
\end{table*}

\section{Results} \label{sec:results}

In this section, we apply the analysis described in Sec. \ref{sec:methods} to the data sample presented in Sec. \ref{sec:data}. In order to follow in detail the different steps of our analysis, we first focus on a single instance (i.e., GW Lup), and then we provide an overview of the global results for the rest of the sample we considered.

\subsection{GW Lup as a case study}
\label{sec:gwlup}

\begin{figure*}
	\centering
	\includegraphics[width=\textwidth]{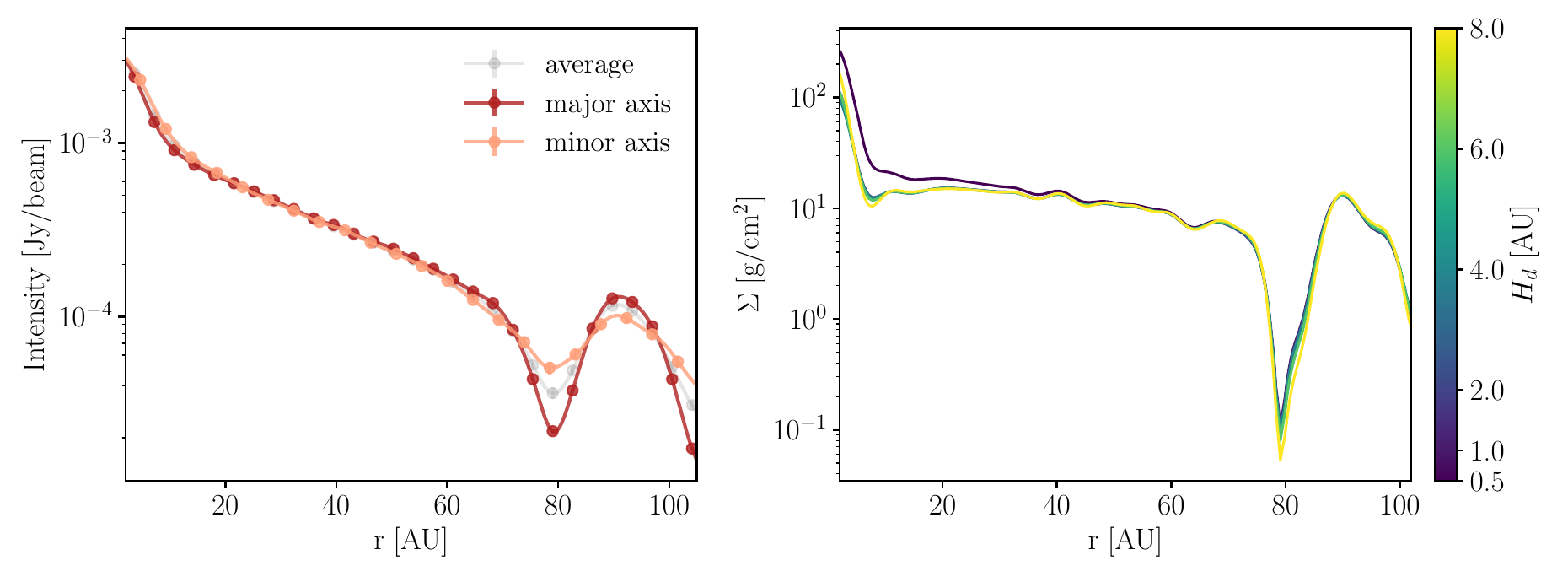}
	\caption{\textit{Left:} Intensity profiles of GW Lup along the minor ($I_{\rm min}^{(\rm data)}$; salmon line) and major ($I_{\rm maj}^{(\rm data)}$; brick) axes, together with the azimuthally averaged profile (gray line). Profiles along the two axes are extracted according to the procedure described in Sec. \ref{sec:iteration}. Errors are computed according to what outlined in Sec. \ref{sec:gwlup}.
	\textit{Right:} Surface density profiles $\Sigma^{(\rm model)}(r;H_d)$ predicted by our model for different disc scale heights $H_d$. The profiles are obtained by using the iteration procedure described in Sec. \ref{sec:iteration}. 
	\label{fig:input_gwlup}
	}
\end{figure*}

GW Lup is a disc with an average inclination of $i\approx40\degree$ and a major structure composed by a gap at $r\approx 74\,\AU$ and a ring at $r\approx85\,\AU$ \citep{DSHARPII}. Because of these properties, it is very well suited for an exemplification of our methodology. 

In the first step of our analysis, we take the observational image and extract the profiles along the major and the minor axes as described in Sec. \ref{sec:methods} (i.e., by averaging the deprojected images in slices that are centred on the axes and have an angular size of $\pi/4$). The resulting profiles are shown in Fig. \ref{fig:input_gwlup} (left panel). We also show the profile obtained by averaging the deprojected image in concentric rings (i.e., the azimuthally-averaged profile). All the different profiles clearly show the characteristic structure of the gap + ring feature. However, as expected, the intensity contrast along the minor axes is smaller due to the \textit{gap-filling effect}.

In order to quantify the statistical uncertainty on the three profiles, we simply compute the standard deviation $\sigma$ of the data in the deprojected images along the slices (or rings) considered, and divide it by the square root of the number of independent data points considered ($\sqrt{N_\mathrm{beams}}$). This latter quantity is simply the azimuthal extent of the slice/ring $\Delta \phi R$, divided by the size of the beam -- which we obtain by averaging the two axes of the beam; along the minor axis, we increase the size of the beam by a factor $\cos i$ to take projection into account. In formula, we get: $\sigma_\mathrm{profile} = \sigma  \sqrt{\mathrm{beam}/\Delta \phi R}$. In Fig. \ref{fig:input_gwlup} (left panel), we plot the error bars only every $N_\mathrm{beams}$, so that they are independent of each other. Note that these errors are very small, and therefore hardly visible in the scale of the plot.

As a second step, we choose a value for the dust scale height parameter $H_d$. In what follows (where not stated otherwise), we consider the following set of values for $H_d$: $\{0.5, 1, 2, 4, 6, 8\}\,\AU$. For each of these values, then, we apply an iteration procedure to match the intensity along the major axis, with the goal of finding the best surface density for the disc (see Sec. \ref{sec:methods} for details on this iteration procedure). The right panel of Figure \ref{fig:input_gwlup} shows the fiducial surface density output by our iteration cycle for different $H_d$. As expected, the predicted surface density is almost identical for different $H_d$ values (with the notable exception of $H_d=0.5\,\AU$).

\begin{figure*}
	\centering
	\includegraphics[width=1.0\textwidth]{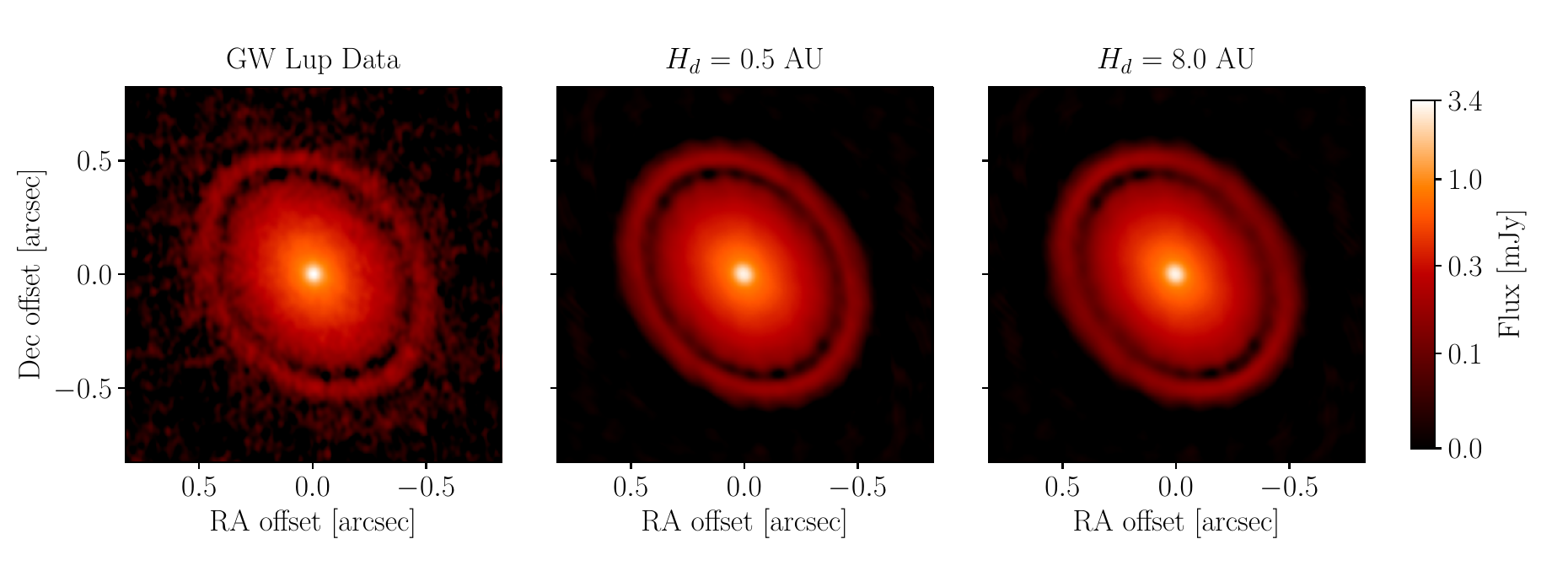}
    \caption{Original image of GW Lup (top left) from the DSHARP survey \citep{Andrews2018}, together with our mock images created using two extreme values of $H_d$ ($H_d= 0.5\,\AU$ and $H_d=8\,\AU$) as well as the surface density profiles shown in Fig. \ref{fig:image_gwlup}. All images are plotted using an $\mathrm{asinh}$ stretch. Mock images are obtained using the same \code{CLEAN} settings as used for the data in DSHARP \citep{Andrews2018}. 
    \label{fig:image_gwlup}
 	}
\end{figure*}

Using these fiducial profiles for the surface density, we can produce mock observations setting the same observational parameters as in Tab. \ref{tab:sample} and using the same configuration as the data (see Sec. \ref{sec:disc_structure} for more details on mock images generation). Figure \ref{fig:image_gwlup} shows these mock images for the two extreme $H_d$ values of $H_d=0.5\,\AU$ and $H_d=8\,\AU$, together with the real observations from DSHARP. Even a quick look at the figures allows us to appreciate how the different systems have a similar intensity along the major axis, whereas they present a different gap filling along the minor one, with the image of the thick disc being significantly more blurred than the one referring to the thin disc.

\begin{figure*}
	\centering
	\includegraphics[width=\textwidth]{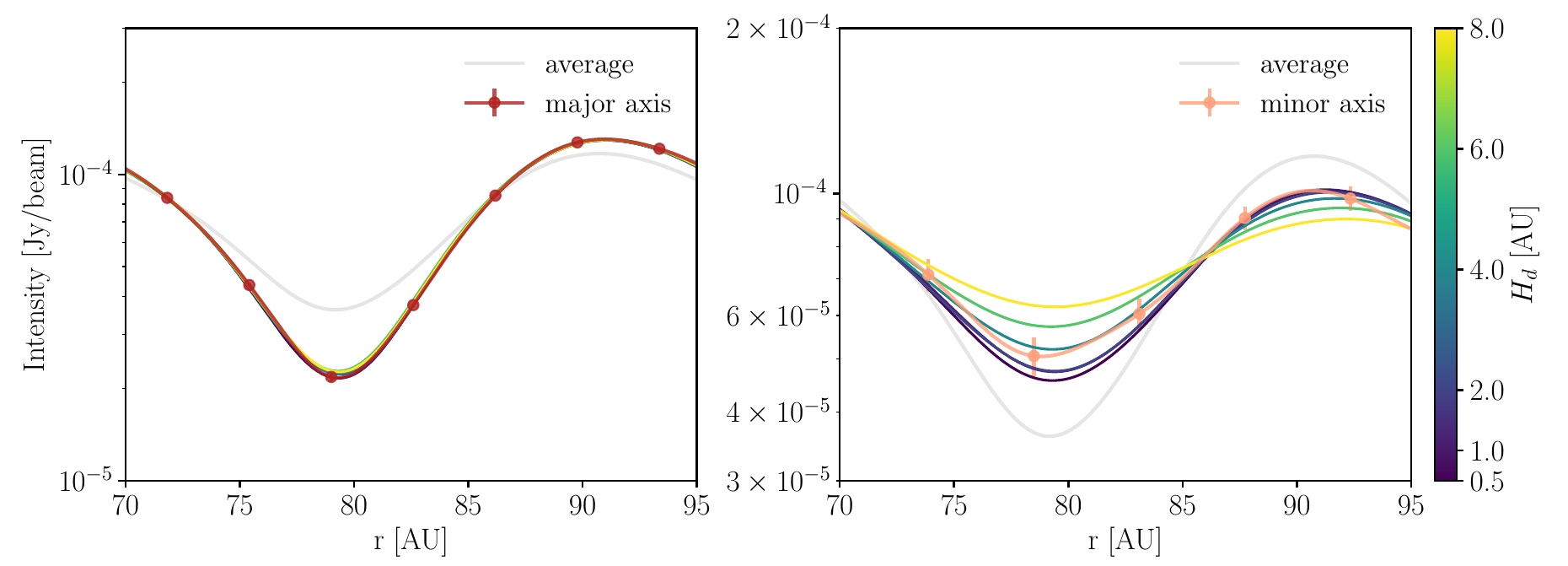}
	
	\caption{\textit{Left:} Comparison of the intensity profiles for GW Lup predicted by our disc model along the major axis (for different values of the disc scale height $H_d$; see color bar) and the one extracted from data (firebrick points). For reference, the observed intensity averaged over the whole azimuthal angle is also plotted with a transparent grey line. The plot only shows a small section of the disc between $r=70\,\AU$ and $r=95\,\AU$, where the major substructures (gap + ring) of GW Lup are present. Predicted intensities align very well with data points, and therefore are almost indistinguishable in the plot.
	\textit{Right}: Same as the left panel, but for the predicted (see color bar) and observed (salmon data points) intensities along the minor axis. Different values of $H_d$ are connected with very different predicted intensities, and this allows us to constrain the true value of $H_d$.
	\label{fig:final_gwlup}
	}
\end{figure*}

This difference can be quantified by deprojecting the images and extracting the profiles along the major and the minor axes in the same way as done with the observational data (i.e., averaging two 1/8-slices of the deprojected images centered along the axes). The resulting profiles for the major (minor) axis are shown in the left (right) panel of Fig. \ref{fig:final_gwlup}, together with the same observational data that are also shown in Fig. \ref{fig:input_gwlup} (left panel). Given that we are interested in the \textit{gap-filling effect}, in the following, we focus only on the region where the gap+ring structure resides (i.e., between $70$ and $95$ AU). 

As expected, the intensity along the major axis is almost the same for any values of the disc scale height $H_d$: all of the different profiles are perfectly compatible with the data. The azimuthally-average intensity from observations is also shown as a reference, in order to highlight how the data vary along different azimuthal axes. The intensity along the minor axis (right panel), on the other hand, strongly varies with $H_d$. In this plot, we can appreciate the predictive power of our method: the \textit{gap-filling effect} implies that for large values of the disc scale height $H_d \gtrsim 6\,\AU$ the resulting profile is much smoother (i.e., the gap is much more filled) with respect to the thin disc cases ($H_d\lesssim 4\,\AU$). Given that the gap in the original data image (salmon data points) is considerably empty, we can conclude that the latter case is to be preferred by observations. Indeed, only the lines with $H_d\lesssim 4\,\AU$ are compatible with the intensity profile of the gap + ring shown by the data. Therefore, we can conclude that the disc GW Lup is thinner than $\approx 4\,\AU$ at $r=100\,\AU$.
In the last columns of Tab. \ref{tab:sample}, we report this conclusion by indicating the constraints we get on the scale height parameter $H_d$.

In order to quantify the agreement between observations and our mock profiles along the minor axis, we choose to employ the $\chi^2$ statistics. However, we caveat that our aim is not to compare models and data in a way that sits on solid statistical bases. This is because, although our iterative procedure works quite well, discrepancies at the level of few percent from the observed emission remain (even along the semi-major axis). These discrepancies are significant given the signal to noise of the observations; in other words, the noise in the data is smaller than our ability to build radiative transfer models that reproduce them. This is a systematic source of error that is not accounted for in a statistics like the $\chi^2$. This does not entail that our method is flawed: in practice, the difference brought upon by the \textit{gap-filling effect} is much larger than the residual discrepancy between data and model. However, given the issues with a detailed comparison between our model and data, we note that the absolute value of $\chi^2$ should not be used to accept or reject models, as it would be the case in a regular statistic test. Nevertheless, for completeness, we report the minimum value of the reduced chi-squared ($\chi^2_{\rm norm}$) in the second to last column of Table \ref{tab:sample}. This is the chi-squared divided by the number of degree of freedom (i.e., the number of independent data points + the number of free parameters in the model). We stress the fact that this number, however, does not have statistical validity and it is not a good parameter to accept/reject our model.

\begin{figure}
	\centering
	\includegraphics[width=0.5\textwidth]{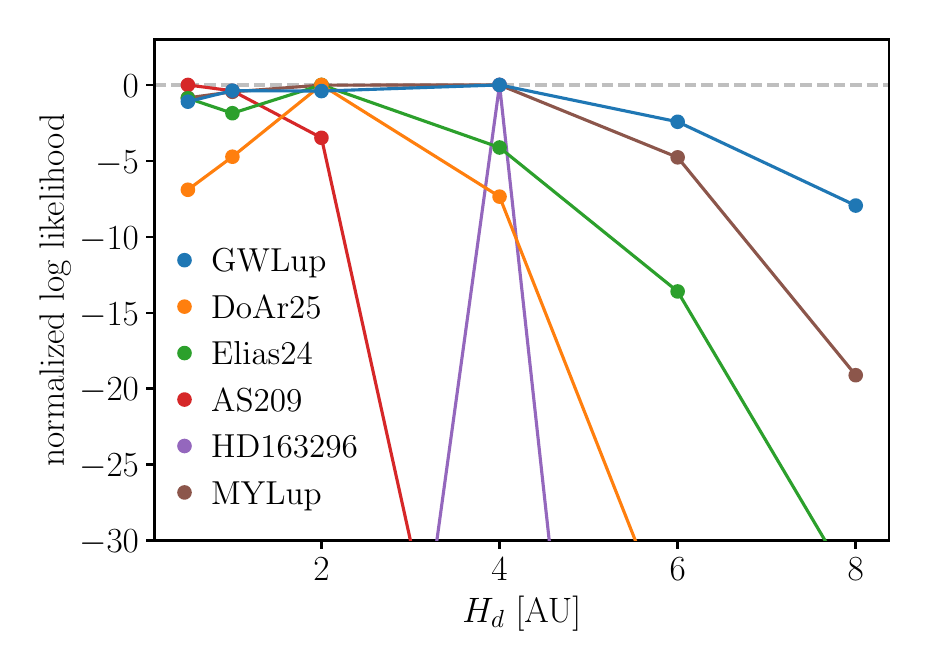}
    \caption{Logarithm of the likelihood ($\log \mathcal{L} \propto -\chi^2/2$) normalised to its peak value, for the values of the parameter $H_d$ considered in our analysis. The systems discussed in Sec. \ref{sec:results} are shown with different colours.
    \label{fig:chi_square}
 	}
\end{figure}

Instead, it is useful to employ the $\chi^2$ as a way to test which of the values of $H_d$ considered in the analysis has a better quantitative agreement with the data. In Fig. \ref{fig:chi_square}, we plot the logarithm of the likelihood function (i.e., $\log \mathcal{L} \propto -\chi^2/2$) normalized to its peak value, for different values of the parameter $H_d$. GW Lup is shown in blue, whereas all the other systems for which we get meaningful constraints on the scale height (see Sec. \ref{sec:other_systems}) are shown in the same plot with different colours. 

From Fig. \ref{fig:chi_square}, we can confirm visually that the best fitting value of the disc scale height is $H_d = 4\,\AU$. However, values that are smaller than $4\,\AU$ are also compatible with data, as the value of the likelihood is smaller but still comparable, especially for $H_d=1\,\AU$ and $H_d=2\,\AU$. Values of $H_d$ greater than $4\,\AU$ have significantly smaller likelihoods, and therefore are rejected by our analysis.

\begin{figure*}
	\centering
	\includegraphics[width=1.\textwidth]{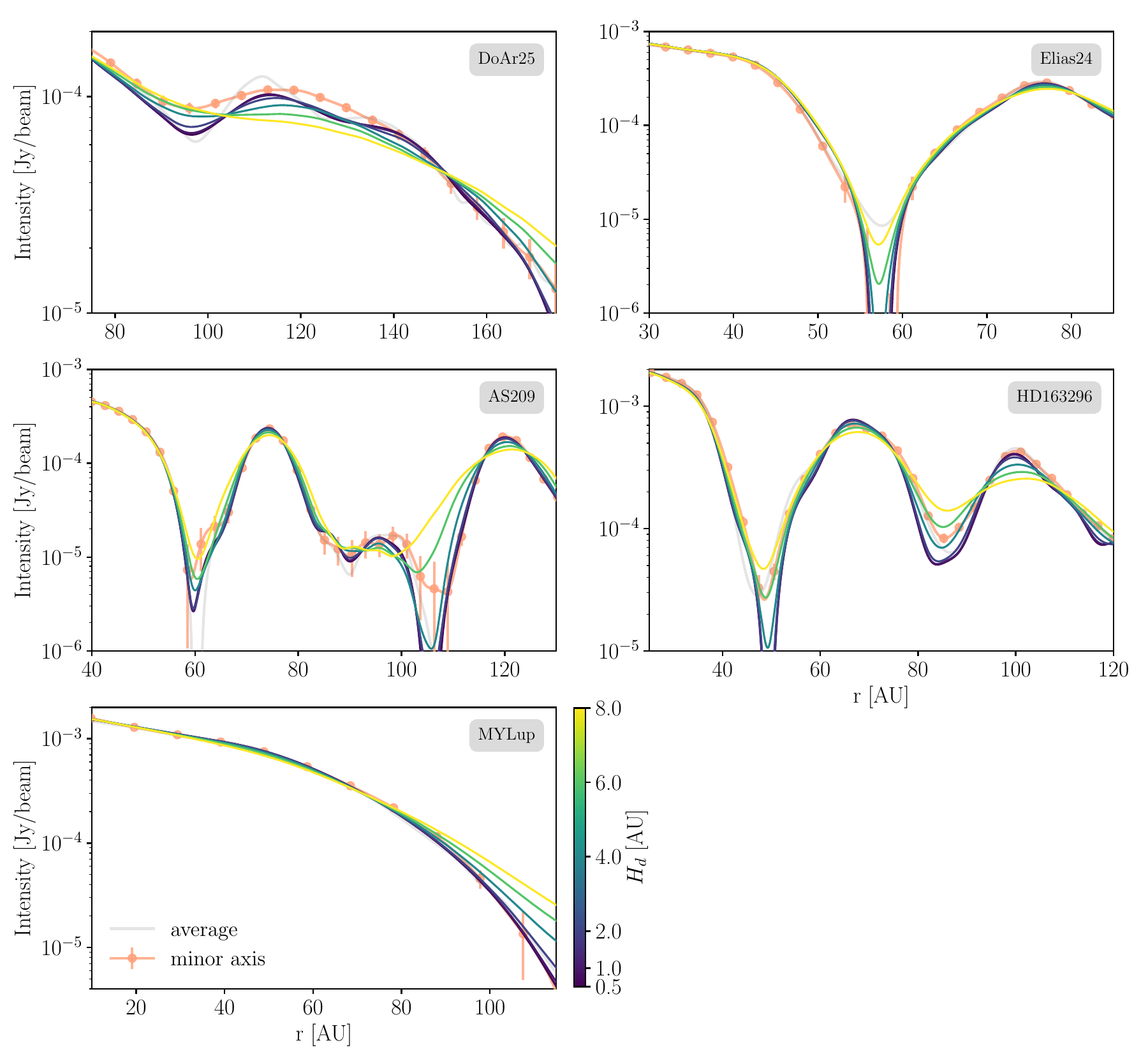}
	
	\caption{Same as the right panel of Fig. \ref{fig:final_gwlup}, but for the systems discussed in Sec. \ref{sec:other_systems}: DoAr 25 (top left), Elias 24 (top right), AS 209 (middle left), HD 163296 (middle right), MY Lup (bottom left).
	\label{fig:final_global}
	}
\end{figure*}

\subsection{Overview of the other systems} \label{sec:other_systems}

In this section, we present the results of our analysis for the remaining systems considered in Sec. \ref{sec:data}. In Tab. \ref{tab:sample} (last columns), we show the constraints we are able to place on the values of the disc scale height based on the comparison between our model and the data along the minor axis. For most of the systems, however, we find that we are unable to place any constraints. This is because the \textit{gap-filling effect} in those systems is not strong enough to produce significant effects on the final predicted intensities. Indeed, we find that for those systems different $H_d$ values produce very similar profiles even along the minor axis. This implies that our approach is not effective in these cases given the structure of the emission and the resolution of observations. We discuss further in Sec. \ref{sec:disc_when_constraints} under which conditions our method is effective in determining the discs' vertical structure.

We choose to focus our discussion in this section only on the systems that yield relevant constraints on $H_d$. The remaining systems -- where our method fails to apply -- are presented in Appendix \ref{sec:appendix_discs_w_no_constraints}. 

Figure \ref{fig:final_global} shows the model intensities along the minor axes for the systems where our method is successful in gauging $H_d$, together with the observed intensities along the same axis (see also the right panel of Fig. \ref{fig:final_gwlup} for the case of GW Lup). In all of these plots (and in the fitting routine), we focus only on the sections of the discs that are relevant for the application of our method and allow us to constrain the value of $H_d$ (e.g., major gaps/rings).

Analogue plots are shown in Appendix \ref{sec:appendix_maj_axis} (Fig. \ref{fig:final_global_major_axis}) for the same systems, but focusing on the major axis instead of the minor one -- same as the left panel of Fig. \ref{fig:final_gwlup}, where we focus on the results of GW Lup only. Intensities along the major axis are generally well recovered by our model because the aim of the convergence procedure described in Sec. \ref{sec:iteration} is to correctly reproduce the observed intensity along this axis. Therefore, this figure serves as a reference in order to test the validity of our approach. 
For completeness, we also include mock images of these systems for $H_d=0.5\,\AU$ and $H_d=8\,\AU$ and compare them with observations in Appendix \ref{sec:appendix_maj_axis}.

In Fig. \ref{fig:chi_square}, instead, we show the $\log$-likelihood as a function of the parameter $H_d$ (normalized to the peak value) for all the systems together. The log-likelihood is computed according to the models and data profiles that are shown in Figure \ref{fig:final_global} (i.e., the intensities along the minor axis).

In the following, we discuss the results of these figures for each system individually.

\subsubsection{DoAr 25}

Due to its large inclination angle ($i\approx67\degree$) and to the presence of a major gap structure, DoAr 25 is a disc where the \textit{gap-filling effect} is quite prominent. Therefore, we expect our method to be effective in discerning which disc scale height is compatible with the observed emission. Indeed, we see (Fig. \ref{fig:final_global}) that different scale heights give rise to very different intensity profiles along the minor axes. However, none of these profiles is perfectly compatible with the observed emission. In fact, the observed gap + ring structure presents an offset with respect to all of the synthetic ones, making it hard to achieve a fair comparison between observations and models. The origin of this offset is unclear; we remark that all the models are converged and can well reproduce the emission profile along the major axis, as can be seen in Fig. \ref{fig:final_global_major_axis}, at least out to $150 \,\AU$ -- beyond which the observations became relatively noisy. The offset may be due to an intrinsic asymmetry in the disc, whereas in our approach we had to assume that the disc is symmetric and any asymmetry is coming from radiative transfer and projection effects \footnote{We have also tried to vary the disc optical depth by changing the normalization of the temperature profile by a factor 2 in either direction, in order to investigate whether optical depth effects could be the cause of the offset. However, we found it to not be the case: results presented here are valid for all the models we experimented with.}. 

Nonetheless, we note that the presence of an observable gap in the minor axis' intensity profile is already a significant probe of a very small disc scale height. This is because, due to the high inclination of the disc, any values of $H_d$ that are $\gtrsim4\,\AU$ would result in an almost complete filling of the gap. Therefore, we conclude that only $H_d\lesssim 2\,\AU$ values are compatible with the observed gap + ring structure along the minor axis. This conclusion is supported by the $\chi^2$ analysis (Fig. \ref{fig:chi_square}), in which we find that the best fitting value of the disc scale height is $H_d=2\,\AU$, with small values significantly preferred to larger ones. 

\subsubsection{Elias 24}

Elias 24 has a very wide and deep gap around $r\approx57\,\AU$. The gap is so deep that, even along the minor axis, the intensity profile presents some negative values. These negatives are due to artefacts created by the CLEAN algorithm; it is notable that they are not present along the major axis and in the averaged profile (gray transparent line). However, given the fact that we adopt the same cleaning procedure as the one used for the data, we can correctly reproduce the profile even when it becomes negative.

Such a deep gap implies, once again, that the disc scale height is very small: only the profiles for $H_d\lesssim 2\,\AU$ show an intensity that becomes negative in the gap centre, whereas larger values of $H_d$ imply at least a partial gap filling along the minor axis and fail to reproduce the CLEANing artefacts. The best $\chi^2$ value, as expected, sits in the range $H_d=0.5-2\,\AU$. 

\subsubsection{AS 209}

The intensity profile of AS 209 is particularly complex: many substructures can be identified both in the inner region of the disc and in the outer one \citep{DSHARPII}. However, only three outer gaps are deep enough to be considered for our analysis of the filling effect. A first major gap is present at $r\approx61\,\AU$, whereas two other gaps $r\approx90\,\AU$ and $r\approx105\,\AU$ form a large, single structure that is delimited by two bright rings at $r\approx74\,\AU$ and $r\approx120\,\AU$, respectively. Therefore, in our analysis, we use this region ($40\,\AU<r<130\,\AU$) to study how the predicted intensities compare with the data. 

Due to the complexity of the observed intensity profile, however, it is hard to tell which profile fits the data better by simply looking at Fig. \ref{fig:final_global}. One thing that is particularly easy to observe is that gaps are fairly deep (and rings fairly bright), and thus very large values of $H_d$ -- represented by green/yellow lines -- are to be excluded. See for example how the green line ($H_d = 6\,\AU$) fails significantly to reproduce the depth of the gap along the minor axis at $105\,\AU$ (Fig. \ref{fig:final_global}), while being a good fit to the major axis (Fig. \ref{fig:final_global_major_axis}).  The $\chi^2$ analysis can quantify this, and it confirms that very small values of the disc scale height ($H\lesssim 2\,\AU$) are preferred over larger ones.

\subsubsection{HD 163296}

HD 163296 is another disc whose morphology is very promising for the application of our method. It has an inclination of $i\approx47 \, \deg$, and two major gaps at $r\approx48\,\AU$ and $r\approx86\,\AU$. The first gap is quite peculiar, as the emission map shows a sizeable blob in the gap along the major axis. This blob represents an issue for our disc modelling, as it is an obvious breaking of azimuthal symmetry. Therefore, we choose to exclude the region containing the blob from our analysis. In order to do that, whenever computing the intensity along the major axis (e.g., to find the surface density with the iteration procedure outlined in Sec. \ref{sec:iteration}), we select only the slice on the side where the blob is not present. We double-check that this choice does not have an influence on the results we find for the outer gap by running a model that includes both sides of the major axis (therefore including the blob, so that the model is only meaningful for the outer gap) and confirming that we obtain very similar emission profiles along the minor axis for the outer gap region.

Looking at the intensity profiles along the minor axis (Fig. \ref{fig:final_global}), we note that there is a broad agreement with data for values of the disc scale height in the range $1\,\AU<H_d<6\,\AU$, depending on the exact gap/ring considered. As a rule of thumb, both gaps are well-fitted by relatively large scale heights ($H_d\approx4-6\,\AU$), whereas the two rings seem to be compatible with lower values of $H_d$. The overall agreement is captured by our $\chi^2$ analysis, which reveals a very strong preference for an intermediate value of the disc scale height ($H_d=4\,\AU$). Therefore, this disc is the only one for which we can place relatively solid constraints on both the upper and the lower limits of the disc thickness. We caveat the reader, however, that the strength of this constraint should be not overestimated. In fact, as also discussed in Sec. \ref{sec:gwlup}, our $\chi^2$ analysis does not take into account the uncertainty associated with our model and relies on some arbitrary assumptions such as the fact that $H_d$ does not vary in different gaps. Indeed, the value of $H_d\approx4\,\AU$ seems to be a compromise between a slightly larger value of $H_d$ in gaps and a smaller value in rings (see Fig. \ref{fig:final_global}). Therefore, we interpret this result as implying that our results do indeed show that HD 163296 is characterized by an intermediate value of $H_d\approx2-6\,\AU$ (and take $H_d\approx4\,\AU$ as our final results), but we do not explore further the exact range of values that are allowed by our $\chi^2$ fitting.

Quite encouragingly, HD 163296 was also analyzed recently by \citet{Liu2022HD163296}. The authors of that study use an analogue method to constrain the vertical structure of the disc, and try to find the best-fitting disc scale height both globally and on every disc/gap separately. In both cases, we see that the values of $H_d$ they find are in broad agreement with the one found here. In particular, we can make a quantitative comparison with their former method, since it is essentially the same as the one used here. Transforming their parametrization of the disc thickness into values of $H_d$ (we do this by assuming a value for the scale height of the gas component, see Sec. \ref{sec:dust_gas_discussion}), they find that the best-fitting profile is the one with $H_d\approx3\,\AU$. This value is very close to the one we find in our analysis.  

\subsubsection{MY Lup}

MY Lup is a very simple disc that does not show any major substructures. The \textit{gap-filling effect} here is thus totally absent. However, the outer edge of the disc is still subject to the same projection effect, and therefore it can be used to determine whether different scale heights produce significant differences in the intensity profile. In other terms, even the outer edge of the disc can be considered part of an "infinitely wide gap" that extends out to infinity starting from the edge of the disc.

Thanks to the high inclination of MY Lup ($i\approx73\degree$), we indeed find that there is a significant difference in the predicted intensity profiles for different values of $H_d$. As shown in Fig. \ref{fig:final_global}, larger $H_d$ values correspond to profiles that are significantly shallower than the observed ones. On the other hand, small scale height ($H_d\lesssim4\,\AU$) profiles present a slope that is generally compatible with data. Therefore, despite the absence of gaps, we can still use MY Lup observations to constrain its disc scale height. 

\section{Discussion} \label{sec:discussion}

In the last section, we applied the method outlined in Sec. \ref{sec:methods} to gauge the dust scale height of DSHARP discs by using the \textit{gap-filling effect}. We have found that: (a) only $\sim 40\%$ of discs yield significant constraints on their dust scale height; (b) for the discs where these constraints are available, we find that the dust scale height (parametrized by $H_d$) is generally low ($H_d\lesssim4\,\AU$), with almost all systems yielding only upper limits to its value. In this section, we discuss the implications of these findings, and we put our results in a broader context by comparing them with previous relevant work on the subject. We conclude by highlighting a few caveats that need to be kept in mind when interpreting our results.   

\subsection{Relative dust and gas scale heights} \label{sec:dust_gas_discussion}

\begin{table}
\centering
\setlength{\extrarowheight}{3pt}
\begin{tabular}{ c | c c c c c }
\toprule
%
 ID  & $H_d$ [AU]    & $H_d/r$   & $\Theta$   & $\alphaSS$/St     & $\alphaSS$ \\ 
 \midrule
\midrule
GW Lup    &    
$\lesssim 4$    &   
$\lesssim 0.04$    &   
$\lesssim 0.44$   &  
$\lesssim 0.24$    &  
$\lesssim 2.4\times 10^{-3}$ \\
DoAr 25    &    
$\lesssim 2$    &   
$\lesssim 0.02$    &   
$\lesssim 0.27$   &  
$\lesssim 0.079$    &  
$\lesssim 7.9\times 10^{-4}$ \\
Elias 24    &    
$\lesssim 2$    &   
$\lesssim 0.02$    &   
$\lesssim 0.20$   &  
$\lesssim 0.042$    &  
$\lesssim 4.2\times 10^{-4}$ \\
AS 209    &    
$\lesssim 2$    &   
$\lesssim 0.02$    &   
$\lesssim 0.25$   &  
$\lesssim 0.065$    &  
$\lesssim 6.5\times 10^{-4}$ \\
HD 163296    &    
$\approx 4$    &   
$\approx 0.04$    &   
$\approx 0.56$   &  
$\approx 0.45$    &  
$\approx 4.5\times 10^{-3}$ \\
MY Lup    &    
$\lesssim 4$    &   
$\lesssim 0.04$    &   
$\lesssim 0.48$   &  
$\lesssim 0.29$    &  
$\lesssim 2.9\times 10^{-3}$ \\
 \bottomrule
\end{tabular}
\caption{ Summary of the constraints on the vertical scale-height and turbulence. Note that, as described in the main text, we have assumed $\mathrm{St}=10^{-2}$ to break the degeneracy between $\alphaSS$ and $\mathrm{St}$.}
\label{tab:turbulence}
\end{table}

As mentioned at the start of Sec. \ref{sec:methods}, we have focused so far only on the dust component of discs because this is the one that can be probed directly by (sub-)mm observations. However, in order to get constraints on the level of turbulence in the disc, we need to consider the vertical structure of the gas component too.

This can be done by assuming that the characteristic value of the gas scale height $h_{g}$ is set by hydrostatic equilibrium ($M_\star$ is the mass of the central star):
\begin{equation}
h_{g} = \sqrt{\frac{k Tr^3}{GM_\star \mu m_p}}
\end{equation}
where $k$ is the Boltzmann's constant, $G$ is the gravitational constant, $m_p$ is the mass of the proton, and $\mu=2.3$ is the mean molecular weight. 

Assuming the gas temperature follows the same relation we already adopted for the dust (eq. \ref{eq:temperature}), then, we can compute the gas temperature everywhere in the disc. With the choice we have made for the radial dependence of the dust scale height (eq \ref{eq:dust_scaleheight_radial}), it can be shown that the gas scale height follows the same dependence, and we can therefore introduce a single parameter $\Theta$, defined as the ratio between the two scale-heights: $h_d(r) = \Theta h_g(r)$. Hereafter, we refer to the parameter $\Theta$ as the \textit{scale height ratio}. 

Given the constraints on $H_d$ we have presented in the previous section, we can use the values of $M_\star$ and $L_\star$ given in Table \ref{tab:sample} and compute the value of $h_g$ at $r=100\,\AU$, and, subsequently, the scale height ratio $\Theta$. We list the resulting values of $\Theta$ in Table \ref{tab:turbulence}. It is easy to note how in all cases the dust scale height is smaller than the gas scale height, as expected from dust settling.

\subsection{Implications for turbulence} 

The ultimate goal of this work is to put constraints on the magnitude of disc turbulence. In order to do this, we follow \citet{Dubrulle1995}, who showed that
\begin{equation}
    \Theta = \left(1+\frac{\mathrm{St}}{\alpha_\mathrm{SS}} \right) ^{-1/2}.
\end{equation}
We list the resulting values of $\alphaSS/\mathrm{St}$ in Table \ref{tab:turbulence}. Note that turning these constraints into a constraint on $\alphaSS$ requires a measurement of $\mathrm{St}$, which at the moment is not available for our whole sample. In the future this may become possible through multi-wavelength observations which measure the spectral index, though significant uncertainties about the dust opacity still remain \citep[e.g.,][]{Sierra2021,Guidi2022}. For the sake of the discussion, we assume here a typical $\mathrm{St} = 10^{-2}$, but we stress this is not a measurement and this is an uncertainty that is carried over to the measurement of $\alphaSS$.

The first thing to note is that all our measurements are incompatible with a value of $\alphaSS=10^{-2}$. This is in line with recent findings in the field that turbulence in proto-planetary discs is relatively weak (see \citealt{RosottiTurbulenceReview} for a review) and also in line with theoretical expectations in the cold conditions of proto-planetary discs, which are not capable to sustain the magneto-rotational instability \citep{BalbusHawley}. For 3 discs, namely half of the sample where we can get constraints, we find even lower upper limits, namely that $\alphaSS < 10^{-3}$, reinforcing the statement that turbulence is weak in proto-planetary discs. Only for one case, HD 163296, our method provides a measurement and not only upper limits, implying that turbulence is (indirectly) detected in this disc. As already discussed, this is in line with the study of \citet{Liu2022HD163296}, who found similar results.

The other aspect we can investigate with our results is whether turbulence is isotropic. In addition to HD163296, which was already discussed by \citet{Liu2022HD163296}, some of our sources have also constraints on turbulence in the \textit{radial} direction: namely AS209 \citep{Rosotti2020}, GW Lup and Elias 24 \citep{Dullemond2018}. Note that these constraints are also obtained by indirectly measuring $\alphaSS$/St. Thus, a comparison between the turbulence level measured in the radial direction and in the vertical one is independent of the assumed Stokes number, St. It is notable that in all three cases the upper limit we derive on $\alphaSS/\mathrm{St}$ is \textit{lower} than the value derived by \citet{Rosotti2020} for AS209 (0.06 with respect to 0.18 and 0.13, depending on which gap/ring we consider) or the lower limit for the range derived by \citet{Dullemond2018} for GW Lup and Elias 24 (0.3 and 0.08, respectively). At face value, this would imply that turbulence in the vertical direction is in fact weaker than in the radial direction. This could have implications regarding the debate on the origin of turbulence, since for example mechanisms like the Vertical Shear Instability (VSI, see \citealt{LesurPPVII} for a review) predict the opposite behaviour because they are particularly effective at lifting particles \citep[e.g.,][]{StollKley2016,Flock2017,Lin2019,Dullemond2022}. Note however that the opposite behaviour is found for HD163296, although the fact it is the only disc in our sample for which we are able to measure the vertical scale height may mean it is exceptionally thick. Considering the small sample size, we are not currently able to draw any conclusions on turbulence anisotropy, but this aspect should be revisited in the future with larger samples. 

\subsection{When does the method yield constraints on the scale height?}
\label{sec:disc_when_constraints}

\begin{figure}
	\centering
	\includegraphics[width=0.5\textwidth]{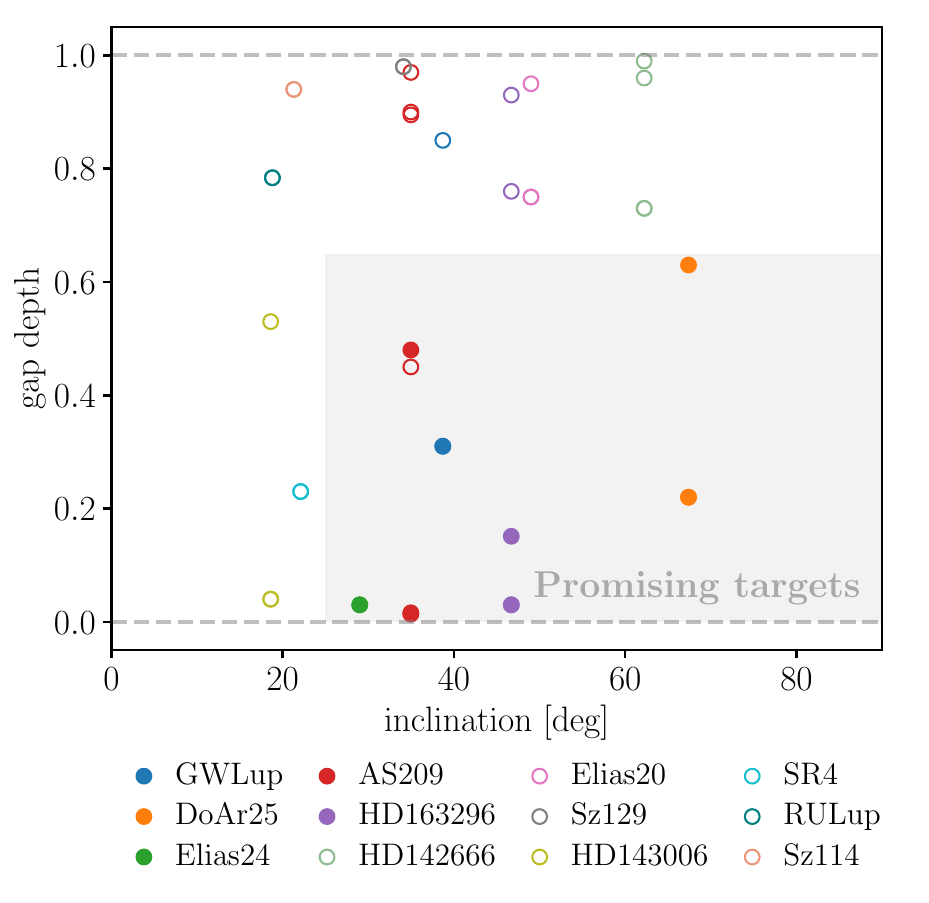}
    \caption{Gap depth (as defined in \citet{DSHARPII}) as a function of the disc inclination angle (as reported in Tab. \ref{tab:sample}) for all the gaps of the DSHARP sample considered in this work for which it is possible to measure a gap depth \citep[for more details, see][]{DSHARPII}. Different colours refer to different systems in the sample. Filled (empty) circles stand for the gap that we (did not) use to measure the dust scale height effectively. The gray shaded region is defined by the two conditions inclination $>25\degree$ and gap depth $<0.65$, and it marks the region where we find that our method proves to be effective in constraining $H_d$.  \label{fig:when_constraints}
 	}
\end{figure}

As we already discussed, for a significant fraction of our discs we are not able to get constraints on the dust scale height. It is worth asking under which conditions the method we use in this paper can give constraints. Considering the method relies on projection effects, we expect it to require discs to have moderate inclinations to be effective. On the other hand, we also expect the method to require deep gaps to work, in order to introduce an appreciable difference between models with different scale heights. On the contrary, shallow gaps are already filled by definition and there is less room for the \textit{gap-filling effect} to introduce a difference between the models. 

In order to quantify more our expectations, we plot in Figure \ref{fig:when_constraints} the properties of gaps in the DSHARP discs listed in Table \ref{tab:sample}. For every gap in these discs, we plot on the x-axis the disc's inclination, and on the y-axis the gap depth taken from \citet{DSHARPII}. This latter quantity is defined as the ratio between the (azimuthally-averaged) intensity in the radial bin containing the centre of the gap and the intensity in the bin containing the centre of the adjacent ring \citep[for more details, see][]{DSHARPII}. If the gap depth could not be measured, we discard the gap from our sample. We use filled (empty) circles to highlight gaps that we (did not) use to effectively constrain the dust scale height. Different systems are plotted using different colours. Note that some of the systems (i.e., AS 209, HD 163296, and GW Lup) have gaps belonging to both of these categories. This is because, in the analysis of these systems (Sec. \ref{sec:results}), we have focused only on the regions where the major (i.e., deeper and larger) gaps reside. Other secondary gaps that were not considered in Sec. \ref{sec:results} are included in Fig. \ref{fig:when_constraints} with empty circles. 

The figure fully confirms our expectations: gaps that can constrain the dust scale height effectively are all residing in a (gray-shaded) region for which $i>25\degree$ and the gap depth is lower than $0.65$. On the other hand, gaps for which our method proves not to be effective are all outside this region, and thus they have either a small inclination or a large gap depth. The sole exception to this is a gap in AS 209 (red empty circle) which has a gap depth of $\sim0.45$ and thus fall in the gray shaded region; however, this gap is located very close to the inner radius ($r\sim9\,\AU$), and therefore it is likely affected by limited spatial resolution. 

We stress that the criterion in which the gray shaded was defined -- although it works well for our sample -- is empirical and should not be taken literally. It is conceivable for example that the specific conditions may vary with the spatial resolution of the observations, as well as with the emission morphology (whose potential variation is presumably larger than what the simple gap depth parameter we introduced can catch). 

Here, we have analysed only the DSHARP sample, as the largest and most homogeneous sample of high-resolution continuum observations. It is unlikely that a single programme will produce a larger sample of high-resolution observations; however, ALMA is conducting more high-resolution campaigns from many programs targeting smaller samples, and combining them one may eventually have a comparable or larger sample than the one we analysed here. The empirical criterion we have derived here may be useful for deciding which ones of those would be worth investigating using the \textit{gap-filling effect}.

\subsection{Caveats} \label{sec:caveat}

The strongest caveat to make regarding this work is that we have implicitly assumed that the disc is azimuthally symmetric. The fact that strong asymmetries are relatively rare is indeed one of the main results of DSHARP \citep{JenningsDSHARP,Andrews2021}, which partially justifies our assumption. We should caveat, however, that here we are interested in rather subtle differences in the azimuthal angle. Therefore, we cannot exclude that asymmetries are indeed present in the discs we observe, but weaker than the obvious ones such as horseshoes, crescents and spirals. This caveat is somehow mitigated by the fact that in the vast majority of cases we can only put upper limits on the dust scale-height, implying that in fact that emission is much more symmetrical (once the different spatial resolution along the major and minor axis is taken into account) than it would be if the disc were thick. The caveat remains however for the example we highlighted of DoAr 25 -- where we are not able to reproduce the emission with an azimuthally symmetric disc -- and for HD 163296 -- where we do measure a scale-height. Though this seems unlikely, we are not able to exclude that the weak asymmetry introduced by the gap-filling effect is instead introduced by an intrinsic asymmetry, and the disc is actually thinner.

Another caveat is that we have taken here a greatly simplified disc temperature structure and we have not set up a realistic grain size distribution. This is done for the sake of simplicity; doing otherwise would introduce many other free parameters regarding the choice of dust opacity, size distribution and disc vertical structure. It is reassuring, though, that for HD163296 our method produces similar results to \citet{Liu2022HD163296}, who did take the more complex route. This is probably because the method we use here is due to projection effects, and as such it should not depend directly on the details of the dust opacity or temperature. 

Finally, we stress the fact that our method can gauge the value of the disc scale height (and hence of $\alphaSS$/St) only locally, where substructures such as gaps and rings reside. Despite the fact that we can reproduce the observed intensity profiles everywhere, it may be that our assumption of a single, global value for the disc scale height does not correspond to reality. In principle, the vertical structure of the disc may vary from one gap to the other one; physical processes such as vortexes at the edge of gaps or meridional flows could also amplify the scale height in the proximity of gaps, biasing the inferred value of the gas turbulence. Therefore, the reader should keep in mind that our conclusions are based on a local effect, and that the knowledge of the behaviour of the scale height globally is an assumption of our model.

\section{Summary} \label{sec:conclusions}

In this work, we have used the \textit{gap-filling effect} to measure the dust scale height in DSHARP discs, with the goal of constraining the amount of turbulence they have. This effect originates from the fact that, in the presence of substructures such as gaps and rings, the intensity profile along the major axis differs from the profile along the minor one. This is because, if the disc inclination is not too small, line-of-sights piercing through the minor axis intercept a larger fraction of the disc's external layers -- which are far from the midplane --, creating a projection effect that ``fills" the gaps along that axis. 

Since this effect is stronger if the disc vertical size is larger, we can probe the value of the disc scale height by building a model whose goal is to reproduce the intensity profiles along the two principal axes. Following previous work by \citet{Pinte2016} and \citet{Liu2022HD163296}, we use radiative transfer to predict the resulting emission maps based on our model. The disc surface density is obtained via an iteration procedure that aims at matching the intensity observed along the major axis. This procedure is successful and convergence is reached at a satisfactory level in almost all cases (see also Fig. \ref{fig:final_global_major_axis}).

The values of the disc scale height ($H_d$; see eq. \ref{eq:dust_scaleheight_radial} for the definition) we find with our analysis can be related to the level of gas turbulence, because the vertical structure of dust grains is set by a competition between gravity and turbulence. Assuming hydrostatic equilibrium for the gas component, we can turn the value of $H_d$ into an estimate for $\alphaSS / $St, and finally into an estimate for the turbulence parameter $\alphaSS$ by assuming a conventional value St $=10^{-2}$ (see Sec. \ref{sec:dust_gas_discussion} for more details).

We summarise here the main findings of this paper: 
\begin{itemize}
    \item We apply our method to 15 discs from the DSHARP survey \citep{Andrews2018}. We manage to successfully constrain the value of disc scale height in 6 of these discs: GW Lup, DoAr 25, Elias 24, AS 209, HD163296, and MY Lup. 
    \item The values of $H_d$ we find are generally very low ($H_d\lesssim 4\,\AU$), and most estimates are upper limits only. In the single case of HD 163296, we can gauge the value of $H_d$ to $H_d\approx4\,\AU$ \citep[in very good agreement with][]{Liu2022HD163296}. 
    \item Turning these values of the disc scale height into constraints for the strength of turbulence (see Table \ref{tab:turbulence}), we find $\alphaSS\lesssim 5\times10^{-3}$. For 3 discs (i.e., half of our sample) we find even lower constraints ($\alphaSS<10^{-3}$). These values are in line with recent findings that suggest a relatively low level of turbulence in protoplanetary discs \citep[for more details, see][]{RosottiTurbulenceReview}.
    \item For the remaining 9 systems in our sample, we find that our method is not effective in constraining the value of the disc scale height: models with very different values of $H_d$ give rise to identical intensity profiles along the minor axes (see Fig. \ref{fig:final_global_no_constraints}). We find that all of these 9 systems ($\approx60\%$ of our sample) are either not very inclined ($i\lesssim 25\degree$) or they host gaps that are not deep enough -- i.e., the intensity at the bottom of the gap is not much smaller than the one in the adjacent ring. We provide an empirical criterion specifying in which region of the inclination-gap depth plane (see Fig. \ref{fig:when_constraints}) the method we employ here can be successfully applied.
\end{itemize}

Looking at the future, the empirical criterion we derive can be used to select from the ever-growing sample (see e.g. the catalogue assembled by \citealt{BaePPVII}) of high-resolution disc observations those where this methodology can be applied, and in this way expand the disc sample with constraints on the vertical scale-height.

Future observations should also focus on gauging the value of the Stokes number. As we have shown in this work, the current sensitivity of observations make it possible to get good constraints on the disc vertical structure (and hence on $\alpha_SS$/St). However, the values we obtain for the level of gas turbulence are subject to our lack of knowledge about the value of the Stokes number, St. Therefore, it is essential in the near future to have complementary multiwavelength observations \citep[see e.g.,][]{CarrascoGonzalez2019,Guidi2022} that can probe the dust grain size distribution - a sub-field that should expand in the next few years thanks to the development of band 1 on ALMA.

\section*{Acknowledgements}

We thank the referee for their comments which improved the clarity of the paper.
EP and GR acknowledge support from the Netherlands Organisation for Scientific Research (NWO, program number 016.Veni.192.233), as this work was originally conceived as part of the LEAPS program at Leiden Observatory. GR also acknowledges support from an STFC Ernest Rutherford Fellowship (grant number ST/T003855/1). B.T. acknowledges support from the Programme National ‘Physique et Chimie du Milieu Interstellaire’ (PCMI) of CNRS/INSU with INC/INP and cofunded by CNES.
This work was funded by the European Union under the European Union’s Horizon Europe Research \& Innovation Programme grant No. 101039651 (DiscEvol). Views and opinions expressed are however those of the author(s) only and do not necessarily reflect those of the European Union or the European Research Council. Neither the European Union nor the granting authority can be held responsible for them.

\section*{Data Availability}

The ALMA data used in this work are available on the ALMA archive at \url{https://almascience.eso.org/aq/}, project code 2016.1.00484.L, and on the website \url{https://almascience.eso.org/almadata/lp/DSHARP/}, as explained in \citet{Andrews2018}. The derived data generated in this research will be shared on reasonable requests to the corresponding author.



\bibliographystyle{mnras}
\bibliography{vert_thickness} 




\appendix

\section{Discs with no constraints} \label{sec:appendix_discs_w_no_constraints}

We show here the results for the discs for which our method is not able to place any constraints on the value of the disc scale height. These are (see also Tab. \ref{tab:sample}): HD 142666, Elias 20, Sz 129, HD 143006, SR 4, RU Lup, Sz 114, WSB 52, DoAr 33. A discussion on why these systems yield no constraints on $H_d$ is made in Sec. \ref{sec:disc_when_constraints}.

\begin{figure*}
	\centering
	\includegraphics[width=1.\textwidth]{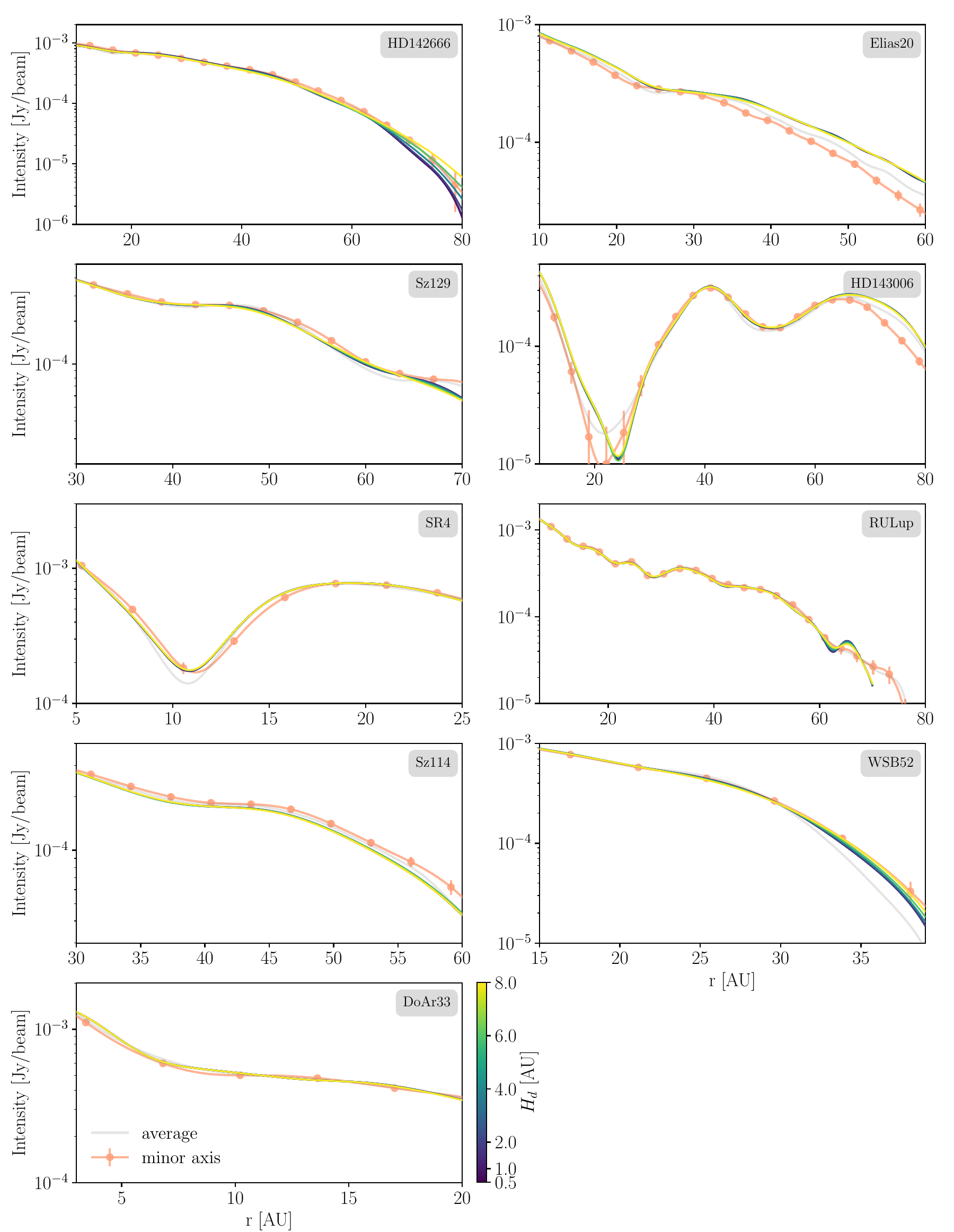}
	
	\caption{Same as Fig. \ref{fig:final_global}, but for the systems of Tab. \ref{tab:sample} for which no constraints on $H_d$ can be placed. \label{fig:final_global_no_constraints}
	}
\end{figure*}

In Fig. \ref{fig:final_global_no_constraints}, we show the intensity profiles along the minor axis extracted from data (salmon lines) together with the ones predicted by our model for different values of the disc scale height $H_d$ (coloured lines). As it is clear from all of the plots, the reason why it is not possible to constrain $H_d$ using our method is that all models with different values of $H_d$ give rise to very similar profiles. 

Thus, despite the fact that these profiles are generally in good agreement with data -- apart from some specific cases where major asymmetries are present, e.g., the outer region of HD 143006 --, we cannot draw any conclusions on the vertical structure of the discs. 

A significant exception to this is the outer region of HD 142666. Similarly to what described in the case of MY Lup (Sec. \ref{sec:other_systems}), profiles with a small (large) value of $H_d$ are much (steeper) shallower due to the same projection effect that takes place in gaps and/or rings. However, in the case of HD 142666, the noise is to high to distinguish which of the different profiles is in better agreement with the data points.

\section{Convergence along the major axis and emission maps}
\label{sec:appendix_maj_axis}

\begin{figure*}
	\centering
	\includegraphics[width=1.\textwidth]{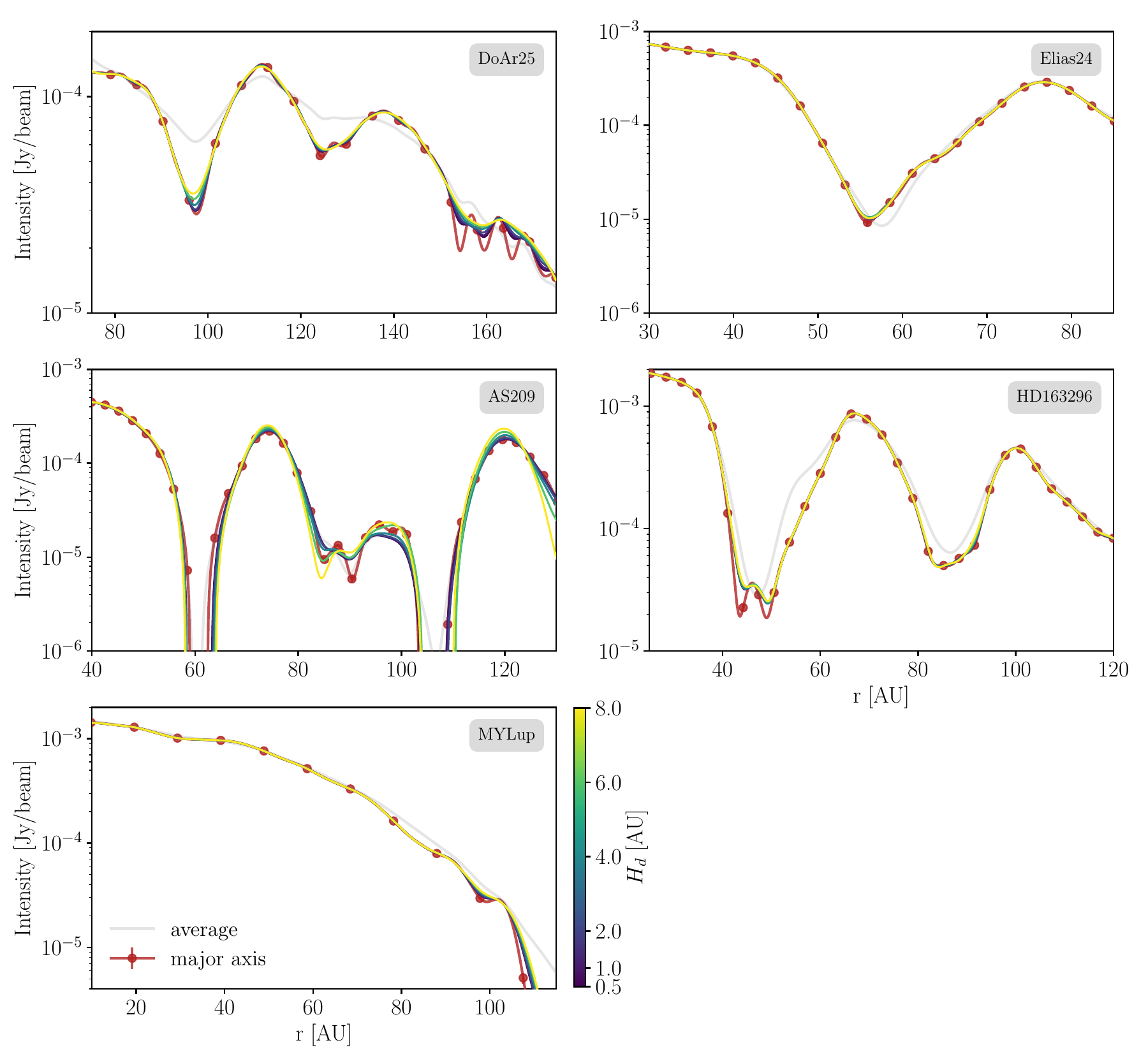}
	
	\caption{Same as the left panel of Fig. \ref{fig:final_gwlup}, but for the systems discussed in Sec. \ref{sec:other_systems}: DoAr 25 (top left), Elias 24 (top right), AS 209 (middle left), HD 163296 (middle right), MY Lup (bottom left). The same figure focusing on intensity profiles along the minor axis is in the main text (Fig. \ref{fig:final_global}).
	\label{fig:final_global_major_axis}
	}
\end{figure*}

\begin{figure*}
	\centering
	\includegraphics[width=1.\textwidth]{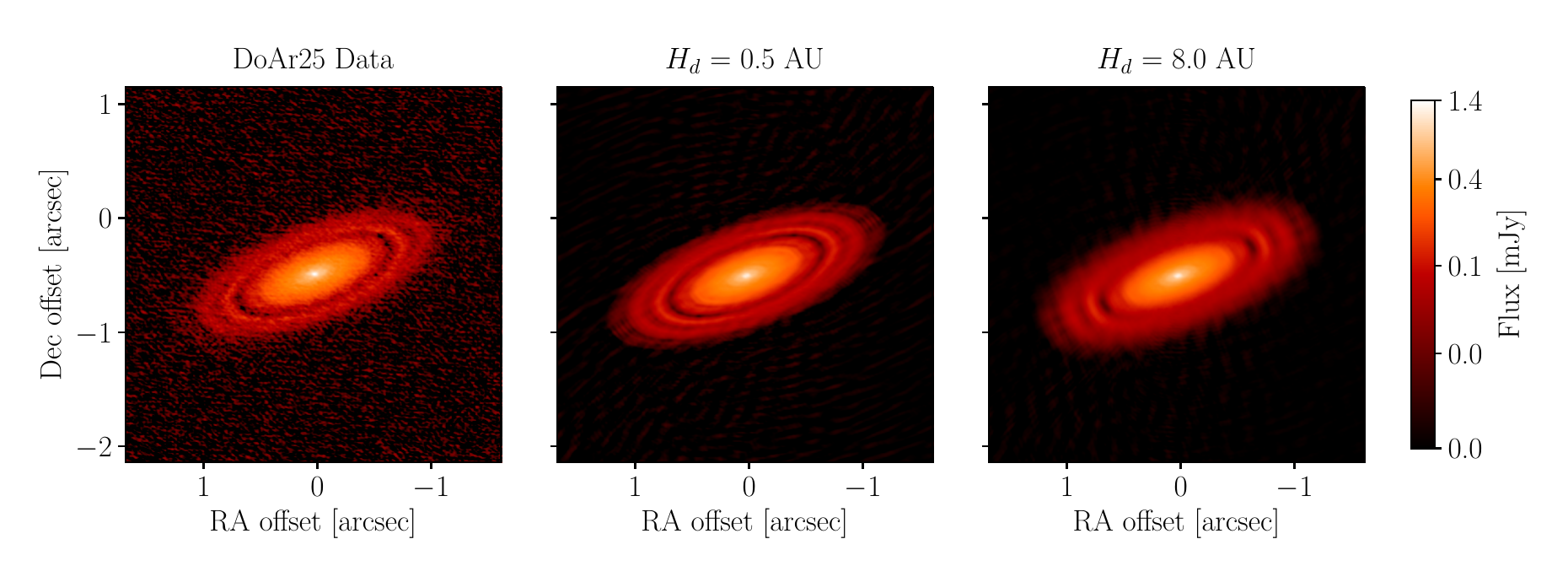}
	\includegraphics[width=1.\textwidth]{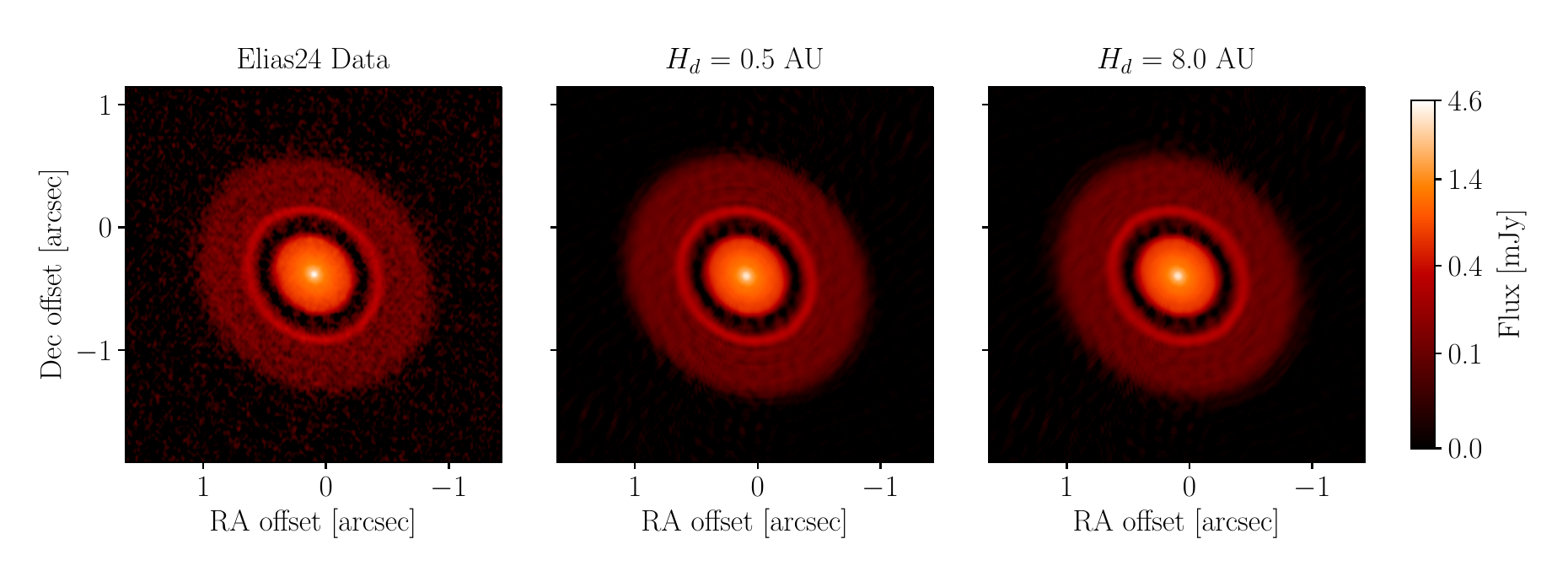}
	\includegraphics[width=1.\textwidth]{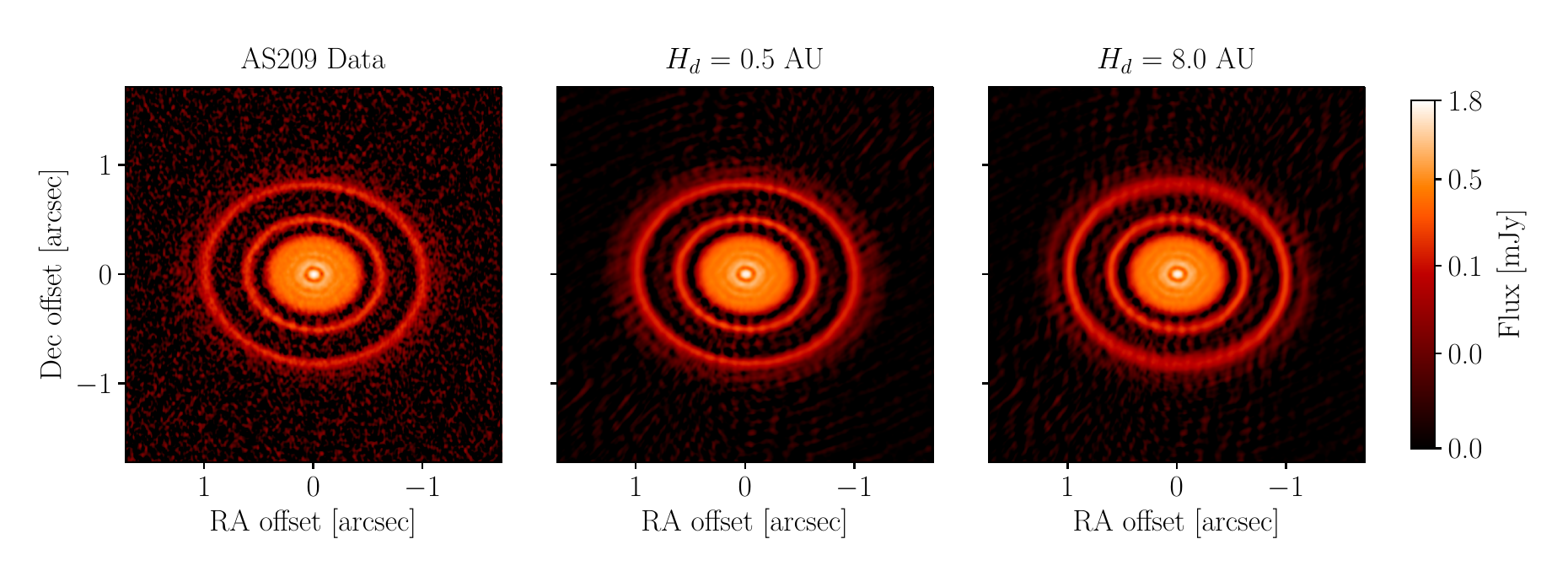}

	\caption{Same as Fig. \ref{fig:image_gwlup}, but for the systems discussed in Sec. \ref{sec:other_systems} (from top to bottom): DoAr 25, Elias 24, and AS 209.
	\label{fig:final_global_mock_images}
	}
\end{figure*}

\begin{figure*}
	\centering

	\includegraphics[width=1.\textwidth]{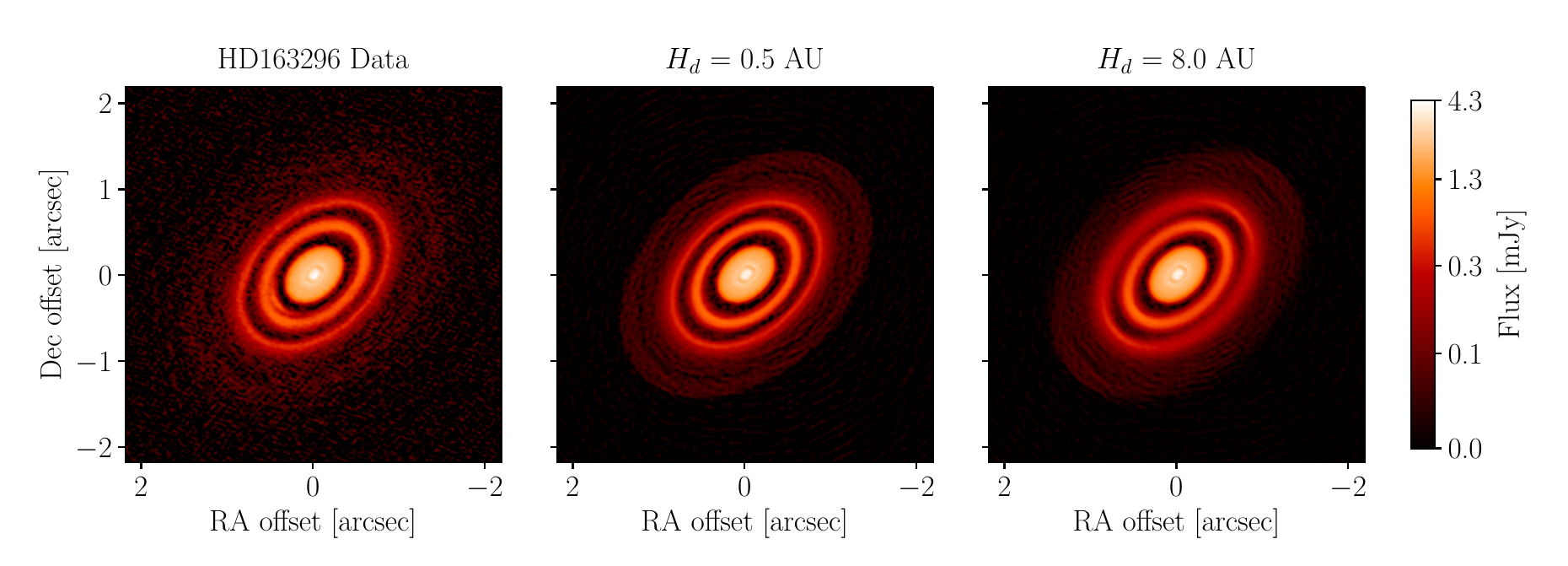}
	\includegraphics[width=1.\textwidth]{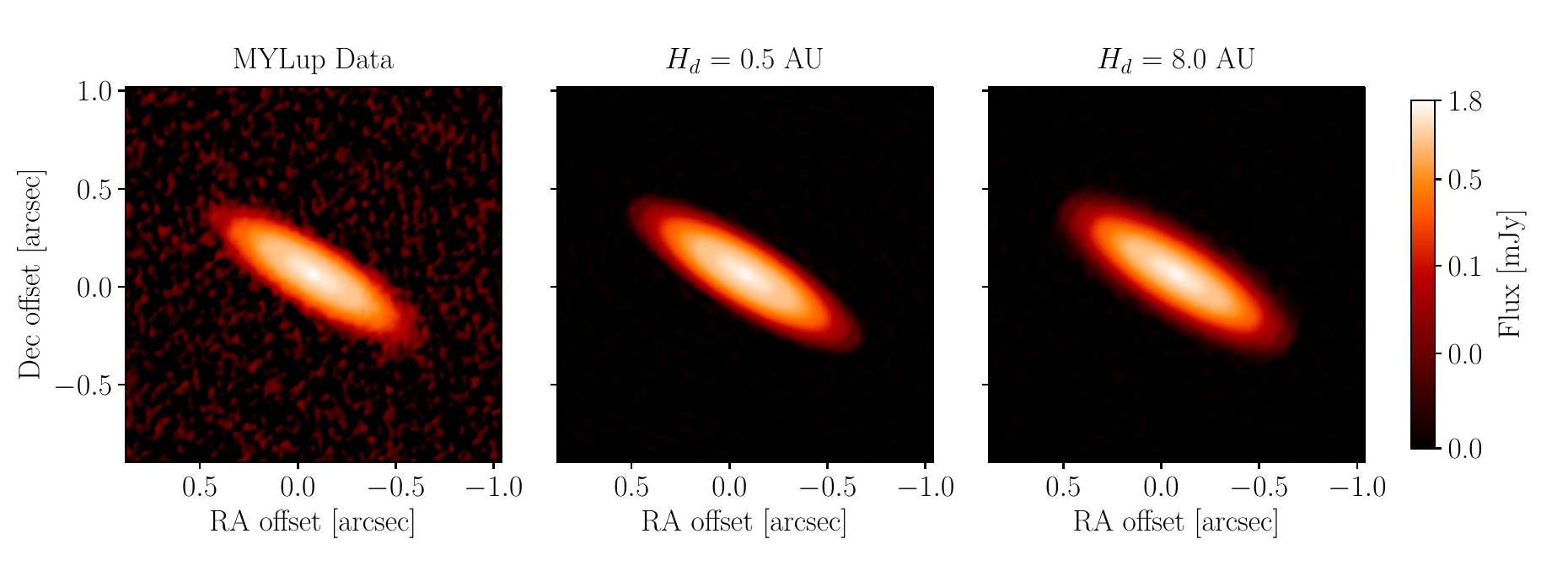}
 
	\caption{Same as Fig. \ref{fig:image_gwlup} and Fig. \ref{fig:final_global_mock_images}, for the remaining systems: HD 163296 (top), MY Lup (bottom).
	\label{fig:final_global_mock_images2}
	}
\end{figure*}

In this section, we show the results of our model-data comparison for what concerns the intensity profiles along the major axis (Figure \ref{fig:final_global_major_axis}) as well as the full mock images of the discs for the two extreme cases $H_d=0.05\,\AU$ and $H_d=8\,\AU$ (Figure \ref{fig:final_global_mock_images}- \ref{fig:final_global_mock_images2}). We focus on the systems that yield significant constraints on the value of the disc scale height (see also Sec. \ref{sec:results} for more details), with the exception of GW Lup which is discussed entirely in the main text (results are in Fig. \ref{fig:final_gwlup}).

\bsp	
\label{lastpage}
\end{document}